\documentclass[longauth]{aa}

\usepackage{graphicx}
\usepackage{txfonts}
\usepackage[dvipsnames]{xcolor}
\usepackage{natbib}
\usepackage{longtable}
\usepackage{siunitx}
\usepackage{multirow}
\usepackage{hyperref}
\usepackage[normalem]{ulem}
\hypersetup{
  colorlinks= true,   
  urlcolor  = blue,     
  linkcolor = blue, 
  citecolor = blue
}

\makeatletter
\renewcommand*\aa@pageof{, page \thepage{} of \pageref*{LastPage}}
\makeatother

\bibpunct{(}{)}{;}{a}{}{,}    %% natbib cite format used by A&A and ApJ

\newcommand{\angstrom}{\mbox{\normalfont\AA}}

\defcitealias{Fotopoulou2012}{FT12}
\defcitealias{Kondapally2021}{K21}

\usepackage{orcidlink}

\begin{document}

\title{The Lockman-SpReSO project}

\subtitle{Description, target selection, observations, and catalogue preparation}

\author{Mauro González-Otero \inst{1,2}\orcidlink{0000-0002-4837-1615}     
      \and
      Carmen P. Padilla-Torres \inst{1,2,3,4} \orcidlink{0000-0001-5475-165X}
      \and
      Jordi Cepa \inst{1,2,4} \orcidlink{0000-0002-6566-724X}
      \and
      J. Jesús González \inst{5}
      \and
      Ángel Bongiovanni\inst{4,6}\orcidlink{0000-0002-3557-3234}
      \and
      Ana María Pérez García\inst{4,7}\orcidlink{0000-0003-1634-3588}
      \and
      J. Ignacio González-Serrano \inst{4,8}\orcidlink{0000-0003-0795-3026}
      \and
      Emilio Alfaro \inst{9}
      \and
      Vladimir Avila-Reese \inst{5}
      \and
      Erika Benítez\inst{5}\orcidlink{0000-0003-1018-2613}
      \and
      Luc Binette \inst{5}
      \and
      Miguel Cerviño \inst{7}\orcidlink{0000-0001-8009-231X} 
      \and
      Irene Cruz-González \inst{8}\orcidlink{0000-0002-2653-1120}
      \and
      José A. de Diego \inst{8}
      \and
      Jesús Gallego\inst{10}\orcidlink{0000-0003-1439-7697}
      \and
      Héctor Hernández-Toledo \inst{5}    
      \and
      Yair Krongold \inst{5}      
      \and
      Maritza A. Lara-López \inst{10}\orcidlink{0000-0001-7327-3489}
      \and
      Jakub Nadolny \inst{11}\orcidlink{0000-0003-1440-9061}
      \and
      Ricardo Pérez-Martínez \inst{4,12}\orcidlink{0000-0002-9127-5522}
      \and
      Mirjana Povi\'c \inst{13,9,14}
      \and
      Miguel Sánchez-Portal \inst{4,6}      
      \and
      Bernabé Cedrés \inst{4}
      \and
      Deborah Dultzin \inst{5}      
      \and
      Elena Jiménez-Bailón \inst{5}
      \and
      Rocío Navarro Martínez \inst{4}
      \and
      C. Alenka Negrete \inst{5}      
      \and
      Irene Pintos-Castro \inst{15}
      \and
      Octavio Valenzuela \inst{5}
      }

\institute{Instituto de Astrofísica de Canarias, 
        E-38205 La Laguna, 
        Tenerife, Spain %1
    \and Departamento de Astrofísica, Universidad de La Laguna (ULL), 
        E-38205 La Laguna, Tenerife, 
        Spain %2
    \and Fundación Galileo Galilei-INAF Rambla José Ana Fernández Pérez, 7,  
        E-38712 Breña Baja, 
        Tenerife, Spain %3
    \and Asociación Astrofísica para la Promoción de la Investigación, Instrumentación y su Desarrollo, ASPID, 
        E-38205 La Laguna, 
        Tenerife, Spain %4
    \and Instituto de Astronomía, Universidad Nacional Autónoma de México, 
        Apdo. Postal 70-264, 04510 
        Ciudad de México, Mexico %5
    \and Institut de Radioastronomie Millimétrique (IRAM), Av. Divina Pastora 7, Núcleo Central 
        E-18012, 
        Granada, Spain %6
    \and Centro de Astrobiología (CSIC/INTA), 
        E-28692 ESAC Campus, Villanueva de la Cañada, 
        Madrid, Spain %7
    \and Instituto de Física de Cantabria (CSIC-Universidad de Cantabria), 
        E-39005, 
        Santander, Spain %8
    \and Instituto de Astrofísica de Andalucía (CSIC), 
        E-18080, 
        Granada, Spain %9
    \and Departamento de Física de la Tierra y Astrofísica, Instituto de Física de Partículas y del Cosmos, IPARCOS. Universidad Complutense de Madrid (UCM), 
        E-28040, 
        Madrid, Spain. %10
    \and Astronomical Observatory Institute, Faculty of Physics, Adam Mickiewicz University, ul.~S{\l}oneczna 36, 
        60-286 Pozna{\'n}, Poland %11
    \and ISDEFE for European Space Astronomy Centre (ESAC)/ESA, P.O. Box 78, 
        E-28690 Villanueva de la Cañada, 
        Madrid, Spain %12
    \and Space Science and Geospatial Institute (SSGI), Entoto Observatory and Research Center (EORC), Astronomy and Astrophysics Research Division, 
        PO Box 33679, 
        Addis Abbaba, Ethiopia %13
    \and  Physics Department, Mbarara University of Science and Technology (MUST), 
        Mbarara, Uganda %14
    \and Centro de Estudios de Física del Cosmos de Aragón (CEFCA), Plaza San Juan 1, 
        44001 Teruel, 
        Spain %15
    \\
    \email{mauro.gonzalez@iac.es, mauromarago@gmail.com}
}        
\date{Received ---; accepted ---}
    
% \abstract{}{}{}{}{}

\abstract
{Extragalactic surveys are a key tool for better understanding the evolution of galaxies. Both deep and wide-field surveys serve to provide a clearer emerging picture of the physical processes that take place in and around galaxies, and to identify which of these processes are the most important in shaping the properties of galaxies.}
{The Lockman Spectroscopic Redshift Survey using Osiris (Lockman-SpReSO) aims to provide one of the most complete optical spectroscopic follow-ups of the far-infrared (FIR) sources detected by the \textit{Herschel} Space Observatory in the Lockman Hole (LH) field. The optical spectroscopic study of the FIR-selected galaxies supplies valuable information about the relation between fundamental FIR and optical parameters, including extinction, star formation rate, and gas metallicity. In this article, we introduce and provide an in-depth description of the Lockman-SpReSO project and of its early results.}
{We selected FIR sources from \textit{Herschel} observations of the central 24 arcmin $\times$ 24 arcmin of the LH field with an optical counterpart up to 24.5 $R_{\rm C}$(AB). The sample comprises 956 \textit{Herschel} FIR sources, plus 188  additional interesting objects in the field. These are point X-ray sources, cataclysmic variable star candidates, high-velocity halo star candidates, radio sources, very red quasi-stellar objects, and optical counterparts of sub-millimetre galaxies. The faint component of the catalogue ($R_{\rm C}(\mathrm{AB})\geq20$) was observed using the OSIRIS instrument on the 10.4 m Gran Telescopio Canarias in multi-object spectroscopy (MOS) mode. The bright component was observed using two multi-fibre spectrographs: the  AF2-WYFFOS at the William Herschel Telescope and the HYDRA instrument at the WYIN telescope.}
{From an input catalogue of 1144 sources, we measured a secure spectroscopic redshift in the range $0.03 \lesssim z \lesssim 4.96$ for 357 sources with at least two identified spectral lines. In addition, for 99 sources that show only  one emission or absorption line, a spectroscopic redshift was postulated based on the line and object properties, and photometric redshift. In both cases, properties of emission and absorption lines were measured. Furthermore, to characterize the sample in more depth with determined spectroscopic redshifts, spectral energy distribution (SED) fits were performed using the CIGALE software. The IR luminosity and the stellar mass estimations for the sample are also presented as a preliminary description.}
{}

\keywords{Astronomical databases: surveys - galaxies: statistics - galaxies: fundamental parameters - techniques: spectroscopic }

\maketitle

\section{Introduction} \label{sec:1}
A fundamental contribution to our understanding of galaxy formation and evolution is the availability of samples of objects as large and as deep as possible. These censuses, or surveys, may be broadly classified as either photometric or spectroscopic.

Photometric surveys observe an area of the sky by integrating a range of wavelengths, commonly referred as photometric bands, pass-bands, or filters. It is common practice to perform observations of the same field in several filters, thus making the spectral coverage as complete as possible. We can differentiate several types of photometric surveys according to the wavelength range integrated during the observations or the resolving power ($R\equiv\lambda/\delta\lambda$, $\delta\lambda$ being the full width at half maximum (FWHM) of the filter's transmission curve) of the used filters. Thus, broad-band surveys have the lowest resolution power ($R\sim5$); the Sloan Digital Sky Survey (SDSS; \citealt{York2000}) and the Dark Energy Spectroscopic Instrument (DESI) Legacy Imaging Surveys \citep{Dey2019} are some examples. In the case of intermediate-band surveys, the power resolution usually ranges between 20 and 40; examples of such surveys are the ALHAMBRA \citep{Moles2008} and the COMBO-17 \citep{Wolf2003} surveys. Narrow-band surveys have the highest resolving power ($R\geq60$); the PAUS survey \citep{Eriksen2019}, the J-PAS survey \citep{Benitez2014}, and the OTELO survey \citep{Bongiovanni2019}, using 40, 54, and 36 narrow-band filters, respectively, are examples.

This photometric information allows us to construct the spectral energy distribution (SED) and fit it using empirical or theoretical spectral templates. A good coverage of photometric information using different filters over a wide spectral range is crucial for obtaining the best possible fits and to accurately determine the physical properties of objects (e.g. stellar mass and IR luminosity), together with their photometric redshift ($z_\mathrm{phot}$).

On the other hand, for each object, spectroscopic surveys yield the spectrum over a wavelength range determined by the instrumental configuration. These spectroscopic surveys generally have brighter flux limits, determined by the achievable resolution. Nevertheless, the multi-object spectroscopy (MOS) mode used in spectroscopic surveys has improved the efficiency of these studies, making it possible to obtain the spectra of tens, hundreds (e.g. the Near-Infrared Spectrograph on the \textit{James Webb Space Telescope}; \citealt{Ferruit2022}), or thousands (e.g. the DESI spectroscopic study; \citealt{Abareshi2022}) of objects simultaneously.
Spectroscopic surveys allow us to obtain more reliable and more accurate spectroscopic redshifts ($z_\mathrm{spec}$) thanks to the possibility of very precise measurements of spectral lines in absorption and emission, both intense and weak, which in turn allow us to determine physical properties of the objects (e.g.\ stellar ages, star formation rate (SFR), extinction, ionization, and gas metallicity).

Numerous spectroscopic surveys of vast numbers of objects have scanned huge areas of the sky; for example, SDSS/BOSS \citep{Dawson2013} has scanned $\sim$10\,000 deg$^2$ in the $i$-band down to 19.9 mag. Other surveys have analysed smaller areas, but in greater depth. For example, the z-COSMOS survey \citep{Lilly2007} conducted studies of the COSMOS field for a total of 30\,000 objects in the redshift range zero to three. The VANDELS ESO public spectroscopic survey \citep{McLure2018} performed a spectroscopic study of sources in the central part of the CANDELS Ultra Deep Survey and the \textit{Chandra} Deep Field South with a redshift range between one and seven. 

Surveys that analyse small areas of the sky in great depth generally tend to do so in sky regions of high Galactic latitude with plenty of broad-band multi-wavelength data. One of these areas that is of great interest is known as the Lockman Hole (hereafter LH) extragalactic field. The LH field is a galactic area with a minimal amount of neutral hydrogen column density ($N_\mathrm{H}$) on the sky \citep{Lockman1986}. This quality makes it one of the best Galactic windows for detecting distant and nearby weak sources, and a perfect target for developing high-sensitivity surveys. The central part of the LH field has a hydrogen column density value of $N_\mathrm{H}=5.8\times 10^{19}\mathrm{cm}^{-2}$ (\citealt{Lockman1986,Dickey1990}). This value is moderately lower than that found at the Galactic poles, where  $N_\mathrm{H}\sim10^{20}\mathrm{cm}^{-2}$ \citep{Dickey1990}. Lower latitudes are unsuitable for this kind of study, given their high extinction.

Because of this exceptional quality, the LH field is of great interest to the scientific community and has been observed over virtually the entire range of the electromagnetic spectrum. In high-energy regimes, missions such as \textit{Chandra}, \textit{XMM-Newton} and \textit{ROSAT} have targeted the field in a quest for deep data per unit observed solid angle. Furthermore, the \textit{GALEX} telescope observed the LH in its two UV photometric bands. Sloan, the Large Binocular Telescope (LBT), Subaru, and UKIRT are examples of telescopes that have observed the LH field at optical wavelengths. In addition, the low-IR background of the LH field (0.38 mJy sr$^{-1}$ at 100 $\mu$m, \citealt{Lonsdale2003}) has prompted deep IR observations such as those carried out with the \textit{Spitzer} and \textit{Herschel} space telescopes. However, despite the wealth of existing data and even though the LH field has been observed from the X-ray to the radio region, there is a surprising lack of deep optical spectroscopic data, which presents an unresolved challenge.

The Lockman Spectroscopic Redshift Survey using OSIRIS (Lockman-SpReSO) aims to address the shortage of spectroscopic information on the LH field by performing a deep spectroscopic follow-up, up to magnitude $R_\mathrm{C}=24.5$, in the optical and near-IR (NIR) ranges, of a sample of galaxies selected from far-IR (FIR) source catalogues. This survey not only determines the spectroscopic redshifts but also estimates the principal properties of the selected galaxies. To this aim, we took advantage of the large collecting surface of the 10 m class Gran Telescopio de Canarias (GTC) telescope and the excellent performance of the MOS mode of the OSIRIS instrument.

This is the first paper of a series that aims to present the Lockman-SpReSO project and it is structured as follows. In Sect. 2, we outline the main features and scientific motivations of the survey. In Sect. 3, we detail the target selection and the development of the source catalogue. In Sect. 4, we describe the planning of the observations and their main properties for the elaboration of the survey. In Sect. 5, we explain how the data reduction was carried out. In Sect. 6, we describe the first results of the spectral line measurements, the determination of the spectroscopic redshift, and the SED-fitting procedure. In Sect. 7, we summarize the main content of the paper and establish a timeline for the next data release of the survey. Throughout the paper, magnitudes in the AB system \citep{Oke1983} are used. The cosmological parameters adopted in this work are: $\Omega_\mathrm{M} = 0.3$, $\Omega_\mathrm{\Lambda} = 0.7$, and $H_\mathrm{0} = 70$ km s$^{-1}$ Mpc$^{-1}$.

%-------------------------------------------------------------------

\section{Lockman-SpReSO} \label{sec:2}

Lockman-SpReSO is a deep optical spectroscopic survey of a sample of mainly FIR-selected objects over the LH field. The region studied is the central $24\times24$ arcmin$^2$ of the LH field with equatorial coordinates (J2000) $10^\mathrm{h} 52^\mathrm{m}43^\mathrm{s}$ $+57\degr 28\arcmin 48\arcsec$ at the centre (north-eastern region).

One of the first studies of optical counterparts of IR sources was that of \cite{Armus1989}, who carried out spectroscopic observations of 53 IR galaxies to determine the nature of the emission from these sources. Another relevant study is that of \cite{Veilleux1995}, who performed a spectroscopic survey of 200 luminous \textit{IRAS} galaxies in order to discern when the IR emission is due to nuclear activity in the galaxy and when it is due to intense starbursts. In their work, they found that the probability of the ionization source coming from nuclear activity increases for cases of higher IR luminosity. More recent studies have consisted of spectroscopic follow-ups of \textit{Spitzer} \citep{Berta2007} and \textit{Herschel} \citep{Casey2012} sources using the Keck observatory. The work by \cite{Berta2007} focused on the study of 35 luminous infrared galaxies (LIRGs) with $z > 1.4$ and has determined that 62$\%$ of the objects with a measurable spectroscopic redshift have an active galactic nucleus (AGN) component. \cite{Casey2012} studied 767 \textit{Herschel} sources by performing a detailed study and estimating luminosity functions for $z < 1.6$. Other studies, using the large statistical database of SDSS, have carried out analyses of IR sources detected with SDSS in the local low-redshift ($z<0.15$) universe (\citealt{Rosario2016}, \citealt{Maragkoudakis2018}).

Many spectroscopic studies have observed the whole LH field in different ranges of the electromagnetic spectrum. From LH observations in the X-ray range made with \textit{XMM-Newton}, a series of papers were published explaining the data obtained \citep{Hasinger2001}, the spectral analysis performed \citep{Mainieri2002}, with a total of 61 spectroscopic redshift identifications, and a catalogue with the fluxes of the sources \citep{Brunner2008}. The \textit{ROSAT} deep survey also published a series of papers studying LH sources: optical identifications, photometry, and spectroscopy with 43 redshifts measured by \cite{Schmidt1998} and 86 by \cite{Lehmann2001}, among others.

\cite{Rovilos2011} made an optical and IR analysis of the properties of AGNs in the LH field detected in the X-ray data described above, and they found 401 optical counterparts to the 409 AGNs detected by \textit{XMM-Newton}. \cite{Patel2011}, using the WYFFOS instrument on the William Herschel Telescope, carried out observations in the optical range of the XMM-LSS and LH-ROSAT X-ray fields and measured a total of 278 and 15 spectroscopic redshifts, respectively. 

Many other studies have performed optical/NIR spectroscopic follow-ups of X-ray sources, given their good quality. \citet{Zappacosta2005}, using the DOLORES instrument in its MOS mode at the Telescopio Nazionale Galileo (TNG), observed 215 sources down to $R=22$ mag, obtained spectroscopic redshifts for 103 objects, and found evidence of a superstructure at $z=0.8$. \citet{Henry2014} postulated that one of the most distant X-ray clusters at $z=1.753$ in the LH field \citep{Henry2010} could actually be a large-scale structure at $z=1.71$. SDSS has also scanned the entire LH field and obtained spectroscopic redshifts for $\sim$115k objects, where only 140 objects down to $r=21.8$ mag lie within the central $24\times24$ arcmin$^2$ of the field \citep{Abdurro2022}. 

Of particular relevance to our work is that carried out by \citet[hereafter FT12]{Fotopoulou2012}, which we describe in more detail in Sect. \ref{sec:3.2}. The authors collected all the available photometric and spectroscopic information on the LH field at the time of publication from the UV (\textit{GALEX}) to NIR (\textit{Spitzer}/IRAC) in a single catalogue, including publicly available good-quality spectroscopic redshifts and the photometric redshifts calculated by themselves.  

More recently, \cite{Kondapally2021} has produced another multi-wavelength catalogue of the radio sources detected by the LOw-Frequency ARray (LOFAR; \citealt{vanHarlen2013}) Two Metre Sky Survey (LoTSS; \citealt{Shimwell2017}), which observed (among other fields) the LH field at 150 MHz down to an RMS of 22 $\mu$Jy beam$^{-1}$. The multi-wavelength catalogue contains photometric information from the UV (\textit{GALEX}) to the NIR (\textit{Spitzer}/IRAC) and identifies the multi-wavelength counterparts to the radio sources detected by LoTSS. Since it is a more recent work than \citetalias{Fotopoulou2012}, the photometric information is more up to date, including measurements in the optical range from the \textit{Spitzer} Adaptation of Red-sequence Cluster Survey\footnote{\url{http://www.faculty.ucr.edu/~gillianw/SpARCS/}} (SpARCS; \citealt{Wilson2009}) and the Red Cluster Sequence Lensing Survey (RCSLenS; \citealt{Hildebrandt2016}). The merging of \citetalias{Fotopoulou2012} and the multi-wavelength catalogue of \cite{Kondapally2021} ensures that we have the most complete multi-wavelength (from UV to NIR) photometric coverage to perform accurate SED fittings (see Sect. 6.3).

Other studies have been carried out at longer wavelengths. \cite{Swinbank2004}, for example, used imaging and NIR spectroscopy to study 30 (four in the LH field) LIRGs pre-selected from sub-millimetre and radio surveys (\citealt{Chapman2003}, \citeyear{Chapman2005}). Another example is the work of \citet{Coppin2010}, who analysed AGN-dominated sub-millimetre galaxy (SMG) candidates using the \textit{Spitzer}/IRAC spectrograph. The northern region of the LH field was studied as part of the SCUBA-2 Cosmology Legacy Survey at 850 $\rm \mu$m and a depth reached at 1$\sigma$ of 1.1 mJy beam$^{-1}$ using the James Clerk Maxwell Telescope (JCMT, \citealt{Geach2017}).

All previous studies have covered different areas and sizes of the LH field for which there were, until the release of SDSS spectroscopic data, $\sim$600 good-quality spectroscopic redshifts for the whole LH field ($\sim$15 deg$^2$). Although the number was significantly increased thanks to the contribution of SDSS, the number of spectroscopic redshifts in the central $24\times24$ arcmin$^2$ has barely increased with $\sim$150 new values but with a limiting magnitude $i \sim 22$.

To address the lack of spectroscopic information and also study specific families of optical counterparts, the main objective of the Lockman-SpReSO project is to obtain deep optical spectroscopy of a selected sample of objects in the LH field to complement the deepest observations of the \textit{XMM-Newton}, \textit{Spitzer} and \textit{Herschel} space telescopes, and radio data \citep{Ciliegi2003}. The primary sample of objects ($> 80\%$ of the total sample) to be observed with the Lockman-SpReSO project consists of sources observed in the \textit{Herschel}/PACS Evolutionary Probe (PEP) programme by \cite{Lutz2011} with robust optical counterparts down to a magnitude of $\sim$ 24.5 in the Cousins $R_\mathrm{C}$ band. They observed the central $24\times24$ arcmin$^2$ region of the field with a depth of 6 mJy (at 5$\sigma$) at 100 and 160 $\rm{\mu}$m within the framework of the time-guaranteed \textit{Herschel}/PACS key project PEP. These observations enabled sampling near the maximum of the SED of active star-forming galaxies (SFG) at high redshifts ($z<3$). This enables the bolometric luminosities to be estimated more accurately. Moreover, the typical spatial resolution of PACS allows us to perform a reliable cross-correlation with optical sources depending on the spectral band. In order to optimize the use of MOS mode observations, the primary catalogue of FIR sources was supplemented with other types of sources in the field under study, as explained in Sect. \ref{sec:3.5}.

Redshifts were obtained by optical spectroscopy using various instruments (OSIRIS, \citealt{JCepa2000}; WYFFOS, \citealt{wyffos2014} and HYDRA\footnote{\url{https://www.wiyn.org/Instruments/wiynhydra.html}}), which are described in Sect. \ref{sec:4}. Spectroscopic observations of selected targets in the NIR domain are planned for the near future.  

Panchromatic studies (from the X-ray to the radio region) of galaxies have been shown to be a valuable strategy for studying the evolution and properties of these objects. The obtained spectra for selected Lockman-SpReSO objects, complemented with ancillary data, were used to derive the stellar masses, SFRs, gas metallicites, and extinctions. Furthermore, using ratios of spectral lines, we were able to separate SFGs from AGNs using a BPT diagram \citep{Baldwin11981, Stasinska2006}. This segregation is also possible using different IR or X-ray emission \citep[and references therein]{Marina2019}. Among other parameters, SED-fitting techniques allow us to estimate stellar masses, while SFRs can be obtained via either FIR luminosities or optical lines \citep{Kennicutt1998}. Gas metallicities can be measured using different optical relations, the aforementioned R$_{23}$ and N2 methods, and also FIR relations \citep[for example]{Pereira2017, Herrera2018}. Finally, FIR over UV luminosities are considered the best method for determining extinctions \citep{Viaene2016}. 

Several studies have shown that LIRGs lie below the mass--metallicity relation for SFGs \citep[and references therein]{Pereira2017}, although the offset could depend on the selection criteria used. However, these studies are limited to local samples. Ideally, they should be extended in order to ascertain whether this relation depends on FIR luminosity or redshift, or on both. Moreover, the study should encompass the possible differences of the fundamental plane of SFG \citep{Maritza2010} for LIRGs and ultra-luminous infrared galaxies (ULIRGs). For example, studies at higher redshifts have shown that spectral lines are more attenuated than the continuum \citep{Buat2020}, whereas \cite{Eales2018}, using \textit{Herschel} data, make a claim for rapid galaxy evolution in the very recent past.

Some of the scientific objectives of the Lockman-SpReSO project, which are developed in forthcoming papers, include studying the possible evolution of the relation of the masses, extinctions, different SFR indicators, and gas metallicities of \textit{Herschel} galaxies with respect to FIR colours, masses, and FIR and radio luminosities.

Compared to other deep spectroscopic surveys, the Lockman-SpReSO project reaches a depth parameter \citep{Djorgovski2012} 1.2 times greater than that achieved by the VVDS/ultra-deep survey \citep{LeFevre2013}. It is also more advantageous in terms of continuum sensitivity and spectral coverage than z-COSMOS \citep{Lilly2007} and AEGIS-DEEP \citep{Davis2007} respectively. These advantages are largely due to the possibility of using the collecting surface of a 10-metre class telescope and a powerful instrument such as OSIRIS. The sensitivities achievable are higher than those attainable by surveys with smaller collecting surfaces (i.e.\ the SDSS survey), even considering the later releases. In addition, our study is, to date, the most complete, extensive and statistically significant of the optical counterparts of the \textit{Herschel} IR sources.

\begin{figure*}
    \includegraphics[width=1.1\textwidth]{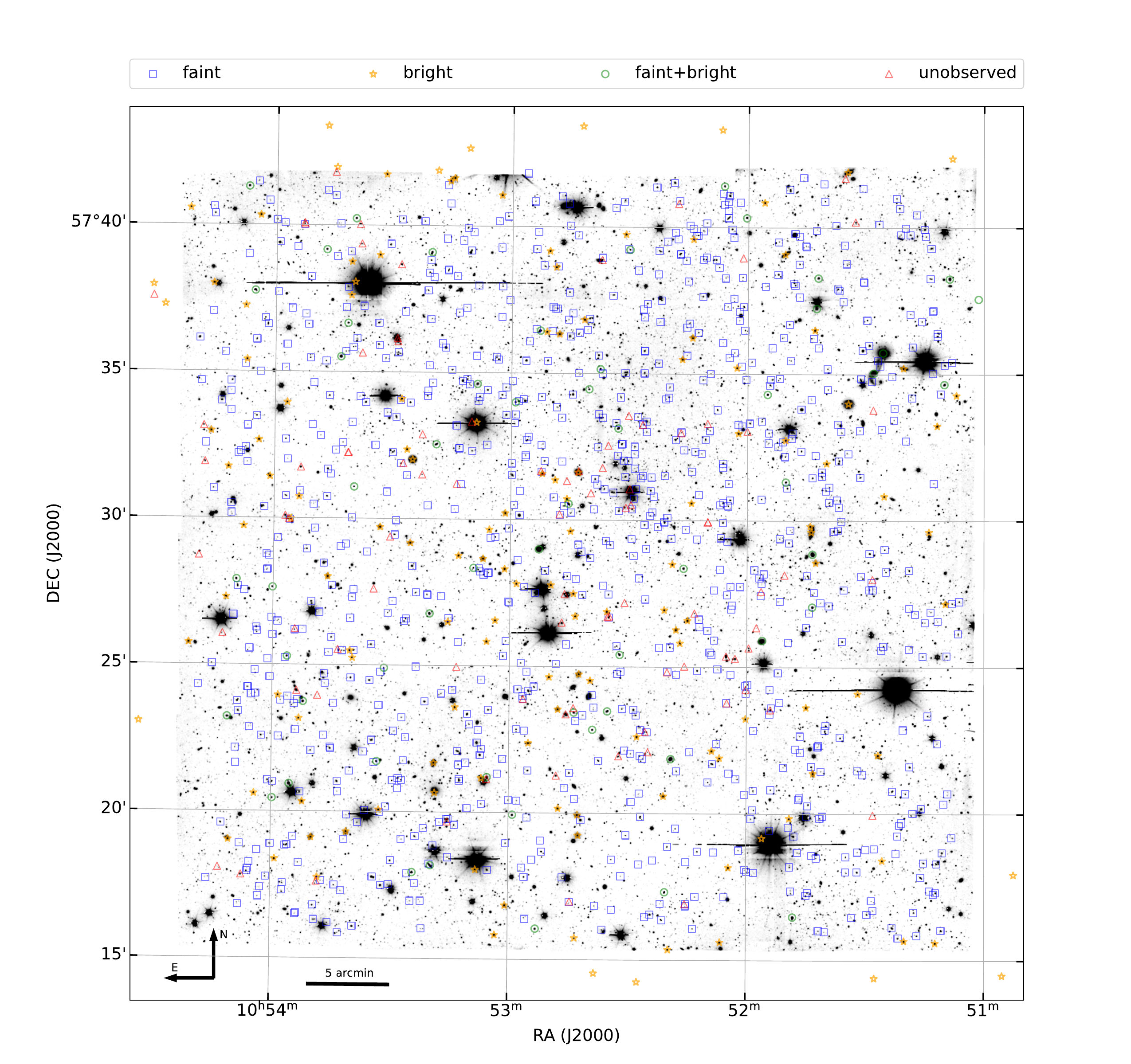}
    \caption{$24\times24$ arcmin$^2$ OSIRIS mosaic (north up, east left) of the studied region of the Lockman-Hole field with equatorial coordinates (J2000) $10^\mathrm{h} 52^\mathrm{m}43^\mathrm{s}$ $+57\degr 28\arcmin 48\arcsec$ at the centre. Objects from the Lockman-SpReSO project input catalogue are highlighted on the mosaic: the blue squares represent objects in the faint subset, the yellow stars are objects in the bright subset, the green circles are objects in both sub-samples, and the red triangles are objects in the catalogue that were left unobserved.}
    \label{fig:mosaic} 
\end{figure*}

%-------------------------------------- 

\section{Target selection} \label{sec:3}
As already mentioned, the main objective of the Lockman-SpReSO project is to provide a high-quality optical spectroscopic follow-up of the FIR sources from the \textit{Herschel}-PEP survey \citep{Lutz2011} with robust optical counterparts in images from OSIRIS in the SDSS $r$ band, up to $R_C=24.5$ mag. This limiting magnitude was originally chosen to reach an S$/$N $\sim 3$ in the continuum with the OSIRIS instrument in MOS mode at resolution $R \sim 500$ and an integration time of about 3 hours, according to the GTC/OSIRIS Exposure Time Calculator\footnote{\url{http://www.gtc.iac.es/instruments/osiris/Osiris_ETC.php}} (ETC, version 2.0).

After selecting the FIR sources and studying which of them have optical counterparts, it was necessary to collect all the good-quality information available in the literature to accurately study the redshifts and physical properties of those sources. Data such as photometric redshifts or magnitudes in various bands are essential for this study.

After an exhaustive search of the literature, three catalogues were used to create the bulk of the sample. The first of these is the catalogue of FIR objects from the PEP Survey \textit{Data Release}\footnote{\url{www.mpe.mpg.de/ir/Research/PEP/public_data_releases}} (DR1) but limited to the central $24\times24$ arcmin$^2$ of the LH field. The second catalogue is that of \citetalias{Fotopoulou2012}, in which the available information from the LH field was collected (see Sect. \ref{sec:3.2}). Finally, the third catalogue was obtained from the broad-band optical images made with OSIRIS (see Sect. \ref{sec:3.3}). The fusion of these three catalogues makes up the primary source catalogue of the Lockman-SpReSO project.

In addition, X-ray emitting counterparts of the FIR sources were identified by using the \textit{XMM-Newton} and \textit{Chandra} mission catalogues. Other secondary catalogues were added to the project to optimize the use of the masks and observation times. The following sections describe the sample selection and creation of the final catalogues.

\subsection{Far-infrared sources} \label{sec:3.1}
As a starting point, we used the data from the DR1 of the PEP survey \cite{Lutz2011} to select the FIR objects. The PEP programme is a \textit{Herschel} guaranteed time extragalactic survey focused on deep PACS \citep{Poglitsch2010} 70, 100, and 160 $\mathrm{\mu m}$ observations of blank fields and lensing clusters. One of them is the LH field, of which complementary observations were made within the HerMES survey \citep{Oliver2010} using the \textit{Herschel}/SPIRE \citep{Griffin2010} photometer and its channels at 250, 350, and 500 $\mathrm{\mu m}$. 

For the purposes of our study, we adopted the catalogue based on the 24 $\mathrm{\mu m}$ priors from \textit{Spitzer}/MIPS, which includes the positions and fluxes as measured by \cite{Egami2008}, and the \textit{Herschel}/PACS photometer fluxes detected by the PEP project at these positions at 100 and 160 $\mathrm{\mu m}$. This information is also available in the published PEP DR1 data. 

After applying the constraint imposed by the coordinates and discarding MIPS sources with no fluxes at 100 and 160 $\mu$m, 1181 sources in total were obtained for the FIR catalogue (hereafter PEP-catalogue).

\subsection{Multi-wavelength catalogue}\label{sec:3.2}
One of the most complete studies carried out on the LH field is that of \citetalias{Fotopoulou2012}, whose authors collected photometric and spectroscopic information available in the literature. They published a catalogue with all the data, including the photometric redshifts for the sources in our field. Specifically, Tables 5 and 10 from \citetalias{Fotopoulou2012} provide the data for photometric information and photometric redshifts respectively. Table 5 in \citetalias{Fotopoulou2012} contains photometric information from the far-UV (FUV) to the mid-IR, in the best case reaching up to 21 bands. Furthermore, spectroscopic redshifts are also included for those objects with high-quality spectroscopic information analysed in 27 studies and compiled in Table 4 of \citetalias{Fotopoulou2012}.

All these data from \citetalias{Fotopoulou2012} were compiled and limited to magnitude $R_{\rm C}\leq24.5$ and coordinates within the $24\times24$ arcmin$^2$ of the field of study. A multi-wavelength catalogue (hereafter the FT-catalogue) of 28\,956 sources was obtained with an astrometric precision better than 0.2 arcsec \citep{Rovilos2011}. The merging of the FT-catalogue and the PEP-catalogue made it possible to limit the FIR sources in $R_{\rm C}$ magnitude.

\subsection{Pre-images of the Lockman Hole field from OSIRIS} \label{sec:3.3}
A total of 64 images were taken to map the central region of the LH field (J2000 equatorial coordinates: $10^\mathrm{h} 52^\mathrm{m}43^\mathrm{s}$ $+57\degr 28\arcmin 48\arcsec$) using OSIRIS in broad-band mode with the SDSS $r$ filter. Those images were meticulously reduced and astrometrically calibrated with RMS $< 0.15$ arcsec. The outcome was a mosaic of the studied field with dimensions of approximately $24\times24$ arcmin$^2$. 

The purpose of the mosaic was to find the optical positions and calibrated magnitudes of the optical counterparts of the FIR sources in the PEP-catalogue in agreement with the imposed flux limit of $R_\mathrm{C}\leq24.5$. The observations were thus designed to achieve a limiting magnitude for the mosaic of $r_{\rm lim} = 25.6$ at $3\sigma$. This limit ensured that all the optical counterparts of the FIR sources were detected. The optical positions determined in this process were those used in the design of the masks for the MOS observation mode.

The extraction of the list of objects and photometric information from the mosaic was implemented using \texttt{SExtractor} \citep{Bertin1996}. A total of 33\,942 sources were extracted with RMS $<0.38$ arcsec, which defines the OSIRIS-catalogue. The resulting mosaic of the OSIRIS pre-image constitutes the background of the map represented in Figure \ref{fig:mosaic}. The fusion of the OSIRIS-catalogue and the PEP-catalogue allowed the IR sample to be limited to those objects with optical counterparts in the OSIRIS mosaic.

\begin{table*}
    \centering
      \caption{Summary of the classes in which the 1144 LH catalogue objects have been detected or for which they are candidates. It should be noted that there are redundancies between the different classes.}
    \label{tab:summary}
    \begin{tabular}{cccccccc}
        \hline \hline
        FIR     & X-ray FIR    & X-ray Point Sources   & High-Velocity & Radio        & Very red   & Sub-millimetre \\
        Sources & counterparts & and Cataclysmic Stars  & Halo Stars    & Sources      & QSOs       & Galaxies      \\
        \hline
        956     & 66           & 58              & 94            & 24           & 70         & 16            \\ 
        \hline 
    \end{tabular}
\end{table*}

\subsection{X-ray counterparts of far-infrared sources}
The X-ray observations in the LH field are the deepest and most complete in the field. These data give us an opportunity to identify the FIR sources with X-ray emission, in addition to optical counterparts up to $R_\mathrm{C}\leq24.5$, which are useful for AGN host classification (e.g.\ \citealt{Povic2009}, \citeyear{Povic2009b}; \citealt{Mahoro2017} and references therein). In particular, we analysed data from the \textit{XMM-Newton}  and \textit{Chandra} missions because of the high quality of their data in the LH field, looking for which sources in the PEP-catalogue have X-ray emission.

The \textit{XMM-Newton} satellite provides the deepest observations in the LH field \citep{Brunner2008}, with a total of 409 sources detected. We matched this catalogue with the PEP-catalogue with a search radius of 2 arcsec, as recommended by  \cite{Povic2009}, and found that 64 were objects in common.

From the \textit{Chandra Source Catalogue} \citep{Evans2010}, we selected the sources in our field under the same conditions as imposed on \textit{XMM-Newton}  data. We identified a total of 106 \textit{Chandra} objects, of which only 19 were matched with the PEP-catalogue, within a search radius of 2 arcsec. 

Finally, we matched both catalogues. We found a total of 66 X-ray sources without redundancies, that is to say 66 FIR sources with emission in the X-ray domain detected by either the \textit{XMM-Newton} or \textit{Chandra} telescopes. All selected objects are also in the FT-catalogue with $R_\mathrm{C}\leq24.5$ mag and in the OSIRIS-catalogue, so the objects are part of the final catalogue as selected objects with IR emission and X-ray counterparts.

\subsection{Additional targets} \label{sec:3.5}
The possibility of working with the GTC's large collecting surface and the efficiency of OSIRIS allows us to complement the Lockman-SpReSO project scientifically with secondary studies. We added interesting complementary targets to take full advantage of the OSIRIS MOS mode and to optimize the design of masks.

Since none of the secondary catalogues impose the criteria of IR emission in their objects, redundancies may arise among the  secondary catalogues, and even with the main catalogue. A study of the existence and correction of redundancies was carried out as the last step in the compilation of the final catalogue of the Lockman-SpReSO project.

Two studies on secondary objects are proposed. For those objects whose nature has been determined in previous studies, further information is expected to be added through optical spectroscopy.  This is the case for galaxies studied in the sub-millimetre and radio domains. On the other hand, for those objects whose nature is indeterminate, spectroscopy helps us reveal the type of object being analysed and its properties, an example being the very red quasi-stellar object (QSO) candidates. The following subsections describe each of the additional catalogues in detail.

\subsubsection{X-ray point sources and cataclysmic variable star candidates}  \label{sec:3.5.1}
A new cross-match was made between the OSIRIS-catalogue and the catalogue of \citetalias{Fotopoulou2012} in order to assign the photometric information, but with a search radius of 5 arcsec. Two different search methods were applied. The first, looking for point sources, imposed the \textit{CLASSSTAR}$>0.95$ constraint and an X-ray detection (a flag in the catalogue of \citetalias{Fotopoulou2012}). The result was a list of 45 objects. The second method was the colour criteria of \cite{Drake2014} ($-0.5<u-g<0.5$ and $-0.5<g-r<0.5$) to identify possible cataclysmic variable (CV) stars. This yielded a total of 21 objects, but eight of these were in common with the point sources. Therefore, this catalogue of secondary objects comprises a total of 58 objects. The spectroscopic study of these objects help us to determine whether or not their nature is stellar, together with a study of the degree of contaminants suffered in the colour-based selection method.

\subsubsection{High-velocity halo star candidates}
Another secondary scientific objective focused on the study of high-velocity halo stars. Considering as a first criterion the stellarity given by the photometry, we selected 94 sources from the \textit{Initial GAIA Catalogue} \citep{Smart2014}  with high proper motion ($>10$ mas/year) and $R_\mathrm{C}>18$ as the sample of stars in the Lockman-SpReSO project.

Our interest was centred on spectroscopically sampling the halo while focusing on stars with high proper motion, which could include Galactic runaway stars. The spectra of the objects may provide rough radial velocity information, but good enough to classify them as halo or disc stars. On the other hand, this sampling could provide us with a better classification between stars, galaxies, and quasars, while indicating the degree of contamination of the stellar sample defined only by the stellarity parameter.

\subsubsection{Radio-source population} 
Deep radio surveys (for more information see \citealt{deZotti2010} and \citealt{Padovani2016}) at levels of a few $\mu $Jy show that there is an excess of sources with respect to the population of powerful radio galaxies. Radio sources above $\sim$1 mJy are typically classical radio sources powered by AGNs and hosted by elliptical galaxies. Below 1 mJy, radio-source counts start to be dominated by SFGs, similar to the nearby starburst population; in other words, radio emission in these galaxies is directly related to the SFR. Thus, deep radio surveys are relevant to the study of the history of star formation in galaxies. One of these surveys in the LH field is the 6 cm (5 GHz) Very Large Array survey by \cite{Ciliegi2003}, who studied 63 radio sources at a depth of $\sim$11 $\mu$Jy. We selected objects from this survey with no spectroscopic information in the bibliography and matched them with the FT-catalogue to add the photometric information. The result was a sample with 24 radio sources with optical counterparts up to $R_\mathrm{C}=24.5$ mag in \citetalias{Fotopoulou2012}.

\subsubsection{Very red quasi-stellar object candidates}
As in the case of the optical spectroscopy of radio galaxies, the scope and possibilities of Lockman-SpReSO give us the opportunity to study other obscured sources that are interesting in their own right. We selected a sample of candidates of very red QSOs by following two different selection processes. The first, proposed by \cite{Glikman2013}, consists in searching for optical, not necessarily point-like objects, counterparts of radio sources from the Faint Images of the Radio Sky at Twenty\footnote{\url{http://sundog.stsci.edu}} (FIRST; \citealt{Becker1995}) survey using the criteria $R-K_{\rm Vega}>4.5$ and $J-K_{\rm Vega}>1.5$. In this way, we found seven very red QSO candidates. 

An alternative selection was based on the work of \cite{Ross2015}, where the colour selection criterion was $r'-W4>7.5$, with $W4$ the 22.19 $\mu$m channel of \textit{Wide-field Infrared Survey} (\textit{WISE}; \citealt{Wright2010}). As a reference to $W4$, they used a relation with MIPS $24 \mu {\rm m}-W4=0.86$ \citep{Brown2014}. For the $r$ band, they used $r'-R_\mathrm{C}=-0.2$ \citep{Ovcharov2008}. In terms of this criterion, 63 sources were classified as potential very red QSOs, yielding a total of 70 by merging both methods.

\subsubsection{Optical counterparts of sub-millimetre galaxies}
As is known, surveys with the \textit{Herschel Space Observatory} have identified an increasing number of SMGs (\citealt{Negrello2010}, \citealt{Mitchell2012}). The study of these sources is of paramount importance for understanding the formation and evolution of massive, dusty galaxies, which could explain the origin of present-day massive ellipticals (e.g. \citealt{Ivison2013}). For these reasons, analysis of its spectroscopic properties in the optical and NIR can be both exciting and challenging, and the Lockman-SpReSO project provides an excellent opportunity to do that at a minimum cost.

The catalogue that was used as a starting point can be found in Table B3 of \cite{Michalowski2012}, which contains the LH field objects detected with JCMT$/$AzTEC  at 1.1 mm as part of the SCUBA HAlf Degree Extragalactic Survey (SHADES; \citealt{Mortier2005}). This survey has a resolution of $\sim$18 arcsec and reaches a depth of $\sim$1 mJy. To determine source identifications, they used the catalogues of \cite{Austermann2010}, thus exploiting deep radio (1.4 GHz) and 24 $\mu$m (0.61 GHz) data, complemented by flux density based methods at 8 $\mu$m and $i-K$ colour. 

To ensure a fair identification of SMG candidates, only objects with very good detection (\textit{bID} $=1$ in Table B1 of \citealt{Michalowski2012}) were selected. We then cross-matched them with those in the PEP-catalogue with a search radius of 5 arcsec and selected those with $R_C\leq24.5$. We obtained a sample of 16 sources with good identification in AzTEC and sub-millimetre emission with optical and FIR counterparts. These SMG sources were also selected as FIR sources with optical counterparts in the OSIRIS mosaic, so they are included in the main object catalogue but flagged regarding that distinctive feature for future studies.

\subsubsection{Fiducial stars}
One of the requirements for quality pointing using the OSIRIS MOS mode is that we have at least three reference points in the field with good accuracy. In this case, these reference points are fiducial stars in the LH field, which help us to point the telescope with accuracy and repeatability. 

Each observation had between three or four fiducial stars to guarantee accurate telescope pointing. Thus, we chose 171 sources in the LH field from SDSS-DR12 \citep{Alam2015}. The sample has a coordinate accuracy of 0.3 arcsec and $R$-band magnitudes between 16 and 19. According to this, the selected fiducials are bright enough to align the mask in a few minutes, but not so bright that they could saturate the acquisition frames when a MOS mask is observed.

\subsection{The final input catalogue}
To compile the final sample, the three main catalogues (the PEP-, FT- and  OSIRIS-catalogues) were merged to obtain a final priority target catalogue. Coordinates from \cite{Egami2008}  were used as reference in the cross-match process because of their high astrometric accuracy. We tested the best value of the maximum allowed distance between the different catalogues. We found that 1.5 arcsec was the best compromise. This is slightly greater than the most significant error of the coordinates in the PEP-catalogue ($\sim$1.22 arcsec).

\begin{figure*}
    \begin{center}
      \includegraphics[trim=1cm 5cm 4cm 0.5cm]{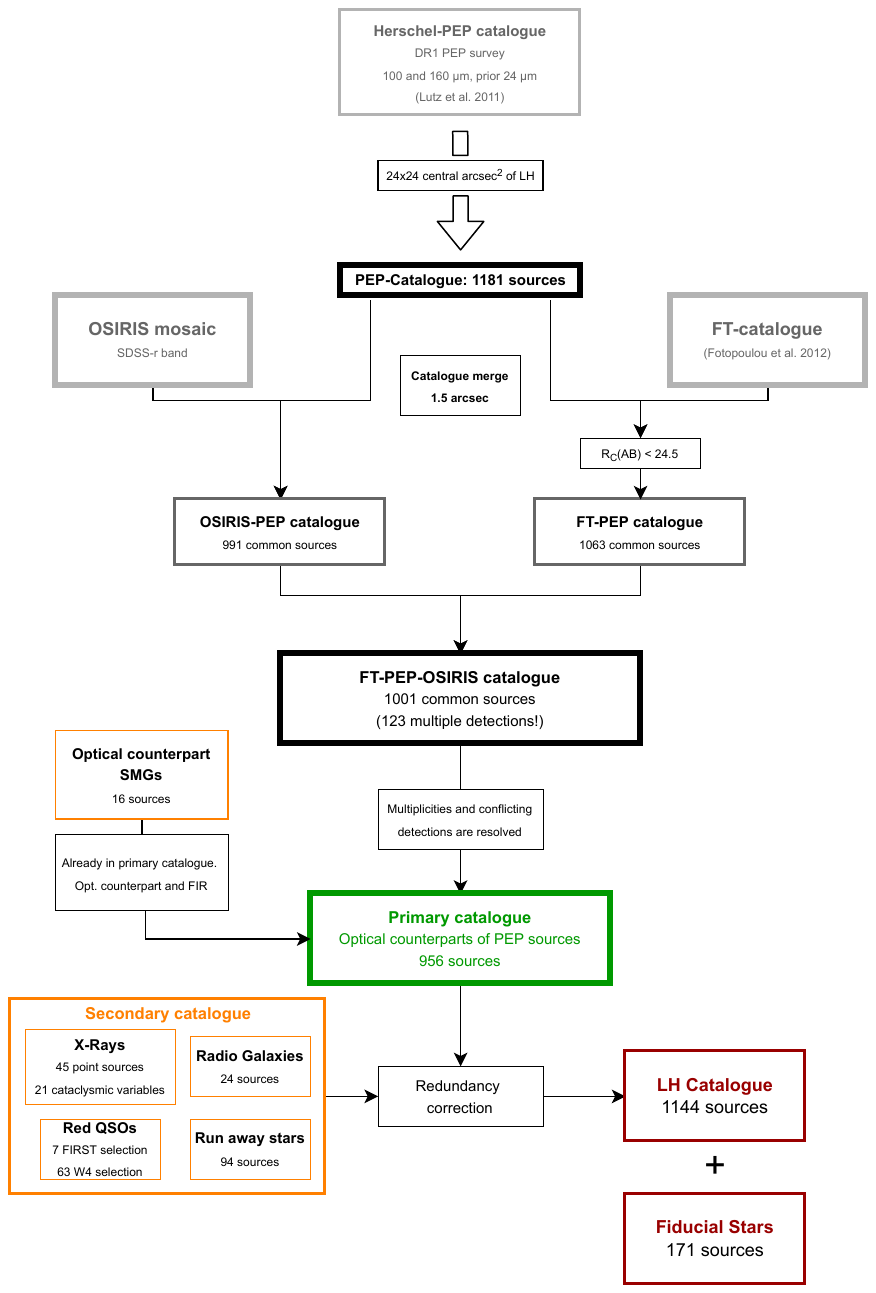}
    \end{center}
    \caption{Merger schedule for the elaboration of the LH-catalogue.} 
  \label{fig:esquema1}
\end{figure*}

We started by joining the target list from the OSIRIS mosaic and the PEP-catalogue. We matched a total of 991 targets within a distance of 1.5 arcsec. After this, we merged the PEP-catalogue with the FT-catalogue to obtain a list of 1063 common objects. The final step was to bring these two previous matches together into a single catalogue. After the correction for multiplicities, the definitive catalogue of primary sources was made up of a total of 956 objects (the primary catalogue). The whole process is described more schematically in Fig. \ref{fig:esquema1} (up to the green box). The last step was to check for possible redundancies between the additional catalogues and the primary catalogue, namely to see if there were objects that appeared in both the primary catalogue and any of the secondary catalogues, while skipping the SMGs that had already been taken into account. The lower half of Fig. \ref{fig:esquema1} represents the merge of the primary and the secondary catalogues. The `Preliminary Object Type' and `Catalogue' columns in Table \ref{tab:z_spec_category} show where the redundancies were found and in what quantity, respectively.

The final target selection from the FIR counterparts in the primary catalogue and complementary sources present in the OSIRIS mosaic, making up the LH-catalogue, includes 1144 sources. The final composition of objects in the LH-catalogue is summarized in Table \ref{tab:summary}. Each value in the table indicates the number of sources in that category that have become part of the LH-catalogue. However, it is important to note that there are redundancies between the different classes; for example, all the SMGs are part of the FIR objects.

%--------------------------------------------------------------------

\section{Spectroscopic observations}\label{sec:4}

\begin{table*}
    \centering
        \caption{Schedule and details of the observing runs over the faint subset observed with OSIRIS/GTC. All the essential information is collected for each run. The number in parentheses in the `Grisms' column represents the number of times that the observations with the grism were made. The first row corresponds to the observation of the images used to elaborate the mosaic.}
        \label{tab:faint_camps}
    \begin{tabular}{ccccccc}
    \hline \hline
    Run     & Masks & OB/Mask   & Grisms    &   Slits Length & Req.\ Night & Exp. Time\\
                          &      (\#)      &            &                      &   (arcsec)            &   & (s) \\
    \hline
    2014A & - & -  & - & -                & - & 10304  \\\hline
    2014B & 10 & 3$^{(1)}$ & R500B(2) \& R1000R(1) & 10 & Gray & 105600  \\
    2015A &  7 & $3^{(2)}$ &  & 10                & Gray & 87400  \\
    2015B &  6 & 3 &  & 10                & Gray & 71400  \\\hline
    2016A &  3 & 3 & R500B(1) \& R500R(2) & 10  & Dark &24000  \\
    2016B &  6 & 3 &  & 10               & Dark & 50400  \\\hline
    2017B & 10 & 4 & R500B(2) \& R500R(2) & 3                 & Dark & 108000  \\
    2018B &  6 & 4 &  & 3                & Dark & 64800  \\
    \hline
    \end{tabular}\\[0.2cm]
    \raggedright\small{$^{(1)}$ Two masks had 2 OBs (R500B(1) and R500R(1)) }\\
    \raggedright\small{$^{(2)}$ One masks had 4 OBs (R500B(2) and R500R(2)) }
    
\end{table*}

Our survey used the guaranteed time of the OSIRIS instrument team and the Instituto de Astronomía of the Universidad Nacional Autónoma de México (IA-UNAM). The first observations were carried out over the first semester of 2014 (run 2014A in Table \ref{tab:faint_camps}) and were used to create the OSIRIS mosaic image of the study region of the LH field (Fig. \ref{fig:mosaic}).

\begin{figure}
    \includegraphics[trim= 0cm 1.5cm 0cm 0cm,width=0.48\textwidth]{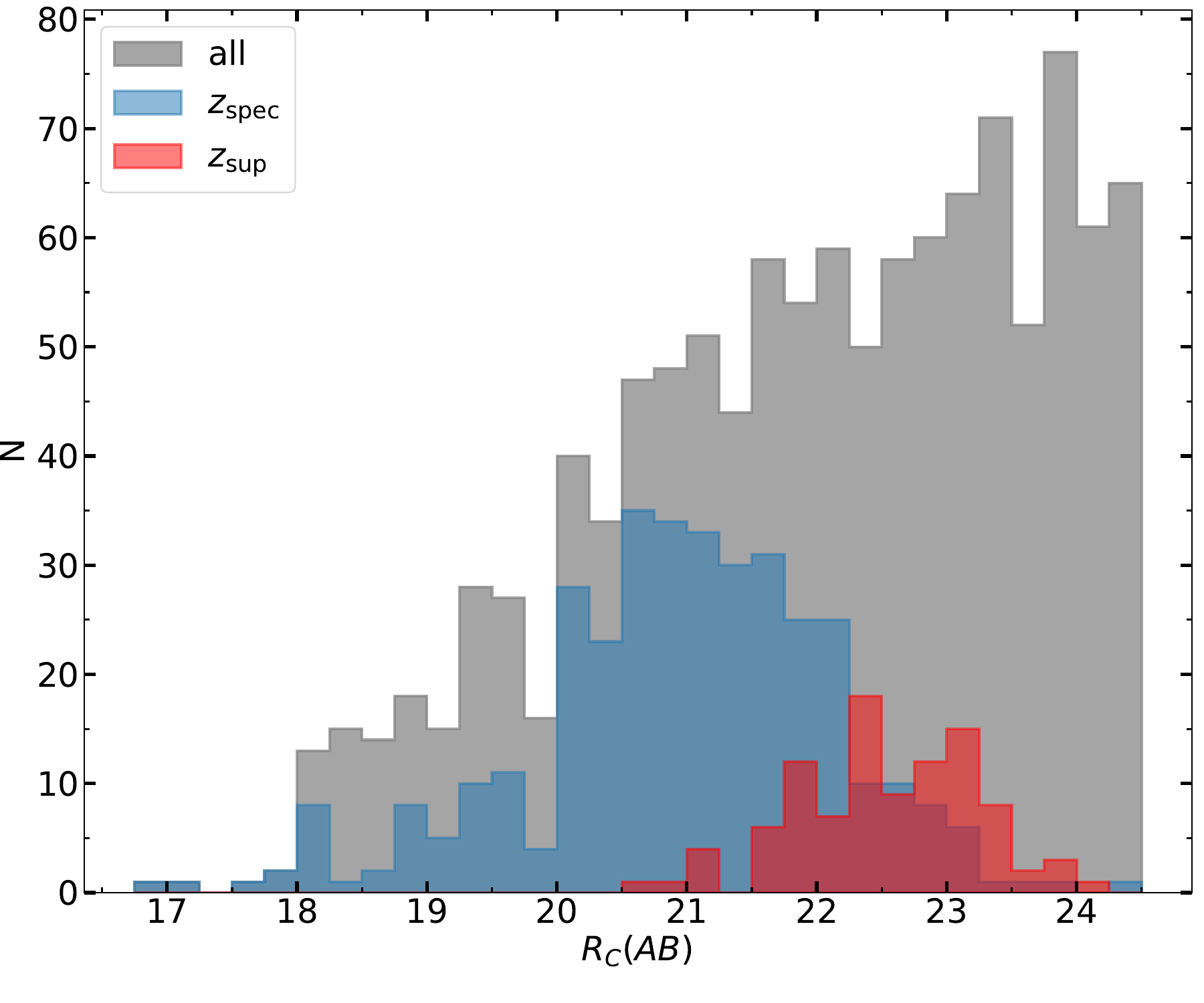}
    \caption{Distribution of $R_\mathrm{C}$ magnitude for the whole Lockman-SpReSO sample (grey), highlighting the objects with $z_\mathrm{spec}$ measured in this study (blue) and objects with $z_\mathrm{sup}$ (red, see Sect. \ref{sec:6.1} for details).}
    \label{fig:mag_Rc}
\end{figure}

As mentioned above, our quality requirement was to reach a S$/$N $\geq 3$ in the continuum for all the objects in the survey. Considering the ETC predictions for the different $R_\mathrm{C}$ magnitude intervals of the survey, as well as the spatial distribution of the objects by $R_\mathrm{C}$ magnitude in the $24\times24$ arcmin$^2$ field, it was determined that, in order to achieve this S$/$N, it would take of the order of 1 to 1.4 h per mask with $R_\mathrm{C}<20.6$ mag, and up to 3 h per mask for sources with $R_\mathrm{C}\geq20.6$ mag. Figure \ref{fig:mag_Rc} shows the $R_\mathrm{C}$ magnitude distribution for all the objects in the catalogue.

Consequently, to avoid problems in merging bright and faint objects, and with the aim of optimizing the number of masks that needed to be used, the sample was divided into two parts using the magnitude criterion $R_\mathrm{C}=20$ mag as the separation value. We denominated objects with $R_C\geq20$ mag as the faint subset (993 sources) and objects with $R_C<20$ mag as the bright subset (151 sources), where each of them was observed with a different telescope. However, there is a slight overlap (up to $R_C=20.6$ mag, 93 sources in the magnitude range 20 $< R_\mathrm{C} <$ 20.6) between the sub-samples for comparing the results obtained with the different telescopes and checking that they are in good agreement. Figure \ref{fig:mosaic} shows the spatial distribution of the objects in each of the subsets, as well as those that are common and those not observed.

\begin{table*}
    \centering
    \caption{Schedule and details of the observations over the bright subset.}
    \label{tab:bright_camps}
    \begin{tabular}{cllccc}
    \hline \hline
    Telescope & Name & Date & Configuration & Grisms & Exp.\ Time \\
              &      &      &               &        &  (sec) \\                 
    \hline
    AF2/WHT    & data 1 & 15/05/2016 & $blue_{1}$ & R600R        & 2$\times$1000\\ 
               & data 2 & 02/06/2016 & $blue_{1}$ & R300B        & 3$\times$1000\\ 
               & data 3 & 19/01/2017 & $blue_{1}$ & R600R        & 3$\times$2000\\
               & data 4 & 20/01/2017 & $red_{1}$  & R600R        & 4$\times$1800+1$\times$1100\\ 
               & data 5 & 05/02/2017 & $red_{1}$  & R300B        & 5$\times$1800 \\ 
               & data 6 & 21/02/2017 & $blue_{1}$ & R300B        & 2$\times$1000 \\ 
               & data 7 & 30/05/2017 & $blue_{2}$ & R600R, R300B & 2$\times$2000,3$\times$1000 \\ \hline
    HYDRA/WYIN & blue1  & 07/05/2018 & $blue_{1}$ & R316R        & 3$\times$1800       \\ 
               & blue2  & 07/05/2018 & $blue_{2}$ & R316R        & 3$\times$1800       \\ 
               & red1   & 08/05/2018 & $red_{1}$ & R316R         & 3$\times$1800       \\ 
               & mixed  & 08/05/2018 & $mix_{1}$ & R316R         & 3$\times$1800       \\ 
    \hline
    \end{tabular}
\end{table*}

\subsection{The faint subset}
The observations of the faint subset were carried out using the OSIRIS instrument at the GTC telescope in MOS-mode.\footnote{\url{www.gtc.iac.es/instruments/osiris/osirisMOS.php}} The first run started in the second semester of 2014 (2014B in Table \ref{tab:faint_camps}), just after the OSIRIS mosaic observations. In addition, spectroscopic MOS mode observations for Lockman-SpReSO also began shortly after the technical commissioning of the OSIRIS MOS mode \citep{Cedillo2018}. The schedule and essential information on the runs can be seen in Table \ref{tab:faint_camps}. The masks were designed using the telescope's software, the OSIRIS Mask Designer   (MD; \citealt{MaskDesigner2004}, \citealt{MaskDesigner2016}). 

Each mask was observed using two different grisms covering the blue and red part of the optical spectrum at an intermediate resolution. The blue region was observed with the R500B grism, which provides a wavelength coverage of 3600--7200 \angstrom\ and a nominal dispersion of 3.54 \angstrom\ pixel$^{-1}$. The red part was covered with two grisms (R500R and R1000R). The former has a wavelength coverage of 4800--10000 \angstrom\ and a dispersion of 4.88 \angstrom\ pixel$^{-1}$, while the latter has a range of 5100--10000 \angstrom\ and a dispersion of 2.62 \angstrom\ pixel$^{-1}$.

The observing strategy changed because of differences in the sample brightness. In this way we could optimize the design of the masks to attain our quality objective of S$/$N $>3$. The first runs were dedicated to observing the brightest objects in the faint subset ($20\lesssim R_C\lesssim 22$, 403 objects), which have a lower density in the field. It was requested that these observations be done on grey nights and the masks were designed using slits with a length of 10 arcsec. This slit length allowed us to select a region in the 2D spectra where the contribution due to emission sky lines could be obtained in order to subtract them from the object spectra (Section \ref{sec:skycorr}). The slit width was set at 1.2 arcsec, as recommended in the official literature for the instrument. This width did not result in a noticeable loss of resolution in the spectrum, and was also in line with the 0.1 arcsec precision of the slit positioning in the OSIRIS MOS mode and the 10$\%$ accuracy requirement in positioning. 

Three observing blocks (OBs) were scheduled for each mask. Two were observed with the R500B grism and one with the R1000R grism. The former had two scientific images per OB, while the latter had three. Table \ref{tab:faint_camps} gives detailed information about the configuration of the runs.

Sky emission subtraction was difficult for faint objects. Moreover, the increase in the number of faint objects in the field implied changing the run's configuration after the 2016B run, when we started to observe the faintest objects in the faint subset ($22\lesssim R_C\leq 24.5$, 535 objects). Each OB was observed in the \textit{ON-OFF-ON} mode (Section \ref{sec:skycorr}), where the \textit{ON-frame} is an actual image and the \textit{OFF-frame} is a consecutive image but with a slight telescope displacement that makes all the slits point to a blank region of the sky. Therefore, each OB had two frames with the spectra of the objects plus sky emission (ON-frames) and one frame with only sky emission spectra (OFF-frame). With this technique, the sky correction for faint objects was greatly improved. Additionally, the sky emission was taken from the \textit{OFF-frame} to enable the slit length to be smaller (3 arcsec) and to allow more slits to be introduced in a given mask design.

The object selection for each mask was made in an effort to minimize the number of masks needed by selecting a field of view with greater object density where possible, but taking care not to merge objects with a very large difference in magnitude and so fit the exposure time in agreement with this. Each mask also had to have some slits for the fiducial stars. These slits were circular with a radius of 2 arcsec. In the masks designed for the faint subset, we also introduced special slits that were pointed to empty regions of the sky in both the \textit{ON-} and \textit{OFF-frames}. These slits were used in the sky correction (see Sect. \ref{sec:skycorr}).

In total, 48 masks, covering 92\% of the objects, were designed to perform observations of the faint subset. All the observations were designed to achieve an S$/$N better than three in order  to obtain good quality spectral lines, even for the faintest objects. In addition to the type of night set for each observation, both airmass and seeing were always requested to be below 1.2. All observations were made in compliance with the required seeing,  60\% with seeing better than 1 arcsec. For the airmass, the constraint was less strict according to the seeing value of that same observation; in other words, if the airmass exceeded the value of 1.2 by never more than 1.35, but the seeing value was good (seeing $<1$ arcsec), the observation was accepted. In total, 12 OBs were earmarked for repetition because the conditions in which they were performed failed to meet requirements, either because of high airmass and seeing or because the type of night was not as requested.

\subsection{The bright subset} 
The bright subset of targets selected for the LH field, $16.8\leq R_\mathrm{C} \leq 20.6$, were observed using two medium-resolution multi-fibre spectrographs, A2F-WYFFOS (AF2) of the WHT at Roque de Los Muchachos Observatory in La Palma, and HYDRA from the WYIN telescope at Kitt Peak Observatory in Tucson.

The selected targets were divided into a `blue sample' ($16.8<r_\mathrm{SDSS}<20.0$) and a `red sample' ($20.0 <r_\mathrm{SDSS}<20.6$). Three fibre configurations were prepared for the blue sample and two for the red sample, to be observed with AF2/WHT. In total, 96 targets were allocated for the blue sample and 50 for the red sample, $\sim60\%$ of the total bright sample plus the overlap with the faint subset, but not all the scheduled observations could be completed.

To complete the observations of the bright subset, we used the HYDRA spectrograph on the WYIN telescope. The sample designed for WYIN was composed of objects not observed with the WHT and some were observed to compare results between telescopes in order to make the most of the observations. Finally, 134 sources with $16.8< R_\mathrm{C}<20.6$ were observed with WYIN.

Regarding the bright part of the catalogue, these objects were expected to be at a lower redshift than in the faint part. Still, as for the faint subset, all observations were set to reach an S$/$N of better than five to obtain quality measurements of the spectral lines.

Figure \ref{fig:mosaic} also shows the spatial distribution of the objects in the bright part. The yellow stars represent the objects in the bright subset, and the green circles represent common objects in both the bright and faint catalogues.

\subsubsection{Multi-fibre optical spectroscopy with AF2-WYFFOS}
We conducted multi-fibre medium-resolution spectroscopy with the AF2 wide-field multi-fibre spectrograph. The AF2 spectrograph contains 150 science fibres of a diameter of 1.6 arcsec, and ten fiducial fibres dedicated to acquiring and tracking guide stars. The AF2 spectrograph has a nominal field of view of 1 deg$^2$, but because of optical distortion and the restriction to avoid the region beyond 25 arcmin, to prevent vignetting of the telescope system and instrument, the configuration file was designed to consider only the area within the central 20 arcmin$^2$. Despite these restrictions, three optimized AF2 pointings were planned to cover the central $24\times24$ arcmin$^2$ of the LH field. With AF2, it is necessary to use the special software \textit{af2-configure}\footnote{\url{https://www.ing.iac.es/Astronomy/instruments/af2/af2_documentation.html}} (from  the Isaac Newton Group) to create  the map of targets to be observed. This software allowed us to optimally place, beforehand, between 50 and 70 fibres per observation on the selected targets. After allocating all the objects in the best way, the remaining fibres were used to observe blank areas in the observation field to obtain sky spectra. 

Observations were performed in three configurations over seven nights in 2016 and 2017 in service mode. More details of the observations are listed in Table \ref{tab:bright_camps}. The R600R and R300B gratings were used with a spectral resolution of 4.4 $\angstrom$  and 3.6 $\angstrom$, respectively. The spectra were centred at wavelength $\sim$5400 $\angstrom$ and covered the range from 3800 to 7000 $\angstrom$, using a 2$\times$2 binning of the CCD camera.

\subsubsection{Multi-fibre optical spectroscopy with HYDRA}
Owing to the decommissioning of AF2 before the planned observations of the bright subset were completed, we had to conduct our observations using a similar instrument. The HYDRA spectrograph on the WYIN telescope was chosen. The Lockman sample for HYDRA had 134 sources with 16.8 $< R_\mathrm{C} \leq$ 20.6. 

The HYDRA spectrograph has 90 active fibres. The targets were observed down to a spectrograph configuration of $\lambda_{\rm start}=4400~\angstrom$, to $\lambda_{\rm end}=9600~\angstrom$, using the the grism $316@7$ with a resolution of $R \sim$ 900 and$\delta\lambda \sim 3~\angstrom$, the lowest resolution but the largest available spectral range.

The images were taken with $3\times 1800$ s exposures for each configuration, adding two series of arcs, and were taken over a total of 8 hours with overheads. We needed four configurations in two half nights. The observations were carried out during the first half of the nights of 8 and 9 May 2018, with a clear sky and a seeing of $\sim$1 arcsec. A summary of the observations is given in Table \ref{tab:bright_camps}.

%--------------------------------------------------------------------

\section{Data reduction}\label{sec:5}
Since the survey data have been obtained using different instruments and telescopes, the nature of the data of each subset is different. Thus, although some procedures are common, the data reduction is described separately for each group of data.

\subsection{Faint subset reduction}
Basic data reduction tasks were carried out using the IRAF-based pipeline GTCMOS (see \citealt{Gomez-Gonzalez2016}) developed by Divakara Mayya of the Instituto Nacional de Astrofísica, Óptica y Electrónica (INAOE), Mexico. For each OB, the first step was to couple the image of the two OSIRIS CCDs into a single one to make it more manageable. We created a master bias and subtracted it from science images, and corrected it for flat-field. The wavelength calibration was performed using Hg, Ar, Xe, and Ne lamp spectra for each grism and mask with a median RMS of $\sim$0.05.

The correction for cosmic rays was carried out using our own python code. To this aim, we relied on the fact that each observation block had at least two science images. This allowed us to compare the same column in both 2D spectra, while looking for pixels more than $3\sigma$ of the median. These pixels are classified as cosmic rays.

\subsubsection{Sky subtraction and flux calibration} \label{sec:skycorr}
The sky emission subtraction of an object's spectrum is both necessary and challenging. The line strength variation of the emission of OH over time is significant, making it painstaking work. As set out above, we performed two different kinds of sky emission subtraction, one for each slit length. 

For the 10-arcsec slits, the subtraction was more direct as we could select regions in the 2D spectra where there was only signal from the sky. To separate the contribution of the observed object in the slit and the sky signal, iterative sigma-clipping was applied column by column. Then, once we obtained the contribution of the sky, we applied a linear fit and finally subtracted it from the original column. The linear fit was implemented to improve the correction because there is a slight curvature in the outer parts of the CCDs  that introduces a distortion in the 2D spectra. The difference between the sky over the column when averaging the sky signal causes the value obtained to diverge from the sky emission and the subtraction fails.

A potential problem with this method occurs when the observed object has a faint continuum because it could be erroneously selected as a sky contribution by the sigma-clipping algorithm, thus resulting in a loss of object information. Furthermore, since the density of faint objects in the field is higher than that of bright objects, the adopted solution was to use 3-arcsec slits observed with the already mentioned \textit{ON-OFF-ON} strategy.

The length limits of the 3-arcsec slits constrained the possibility of selecting sky signal in the 2D spectra. Observing with the \textit{ON-OFF-ON} strategy, we had available direct sky emission spectra (OFF-frame) that could be subtracted directly from the images of the objects (ON-frames). However, the residuals obtained with direct subtraction are considerable. This difference in the sky signal between two consecutive frames is due to the significant variability of the sky emission over the time of the exposures. The solution to this problem was to introduce slits pointing to a region without objects (sky-slits), even in the ON-frames, in such a way that the sky-slits collected sky emission in both ON- and OFF-frames. Thus, if we compare the sky spectra obtained by the sky-slit in one of the ON-frames with the spectra obtained by the same slit,  the result in the OFF-frame is a matrix of sky variation coefficients that can be applied to the OFF-frame spectra to correct the variation of the sky emission over time. Each mask should have at least one sky-slit per OSIRIS CCD to deal with the existing spatial variation of the sky emission, in addition to the time variation.

\begin{figure}
    \begin{center}
        \includegraphics[trim= 0cm 0.4cm 0cm 0cm, width=0.49\textwidth]{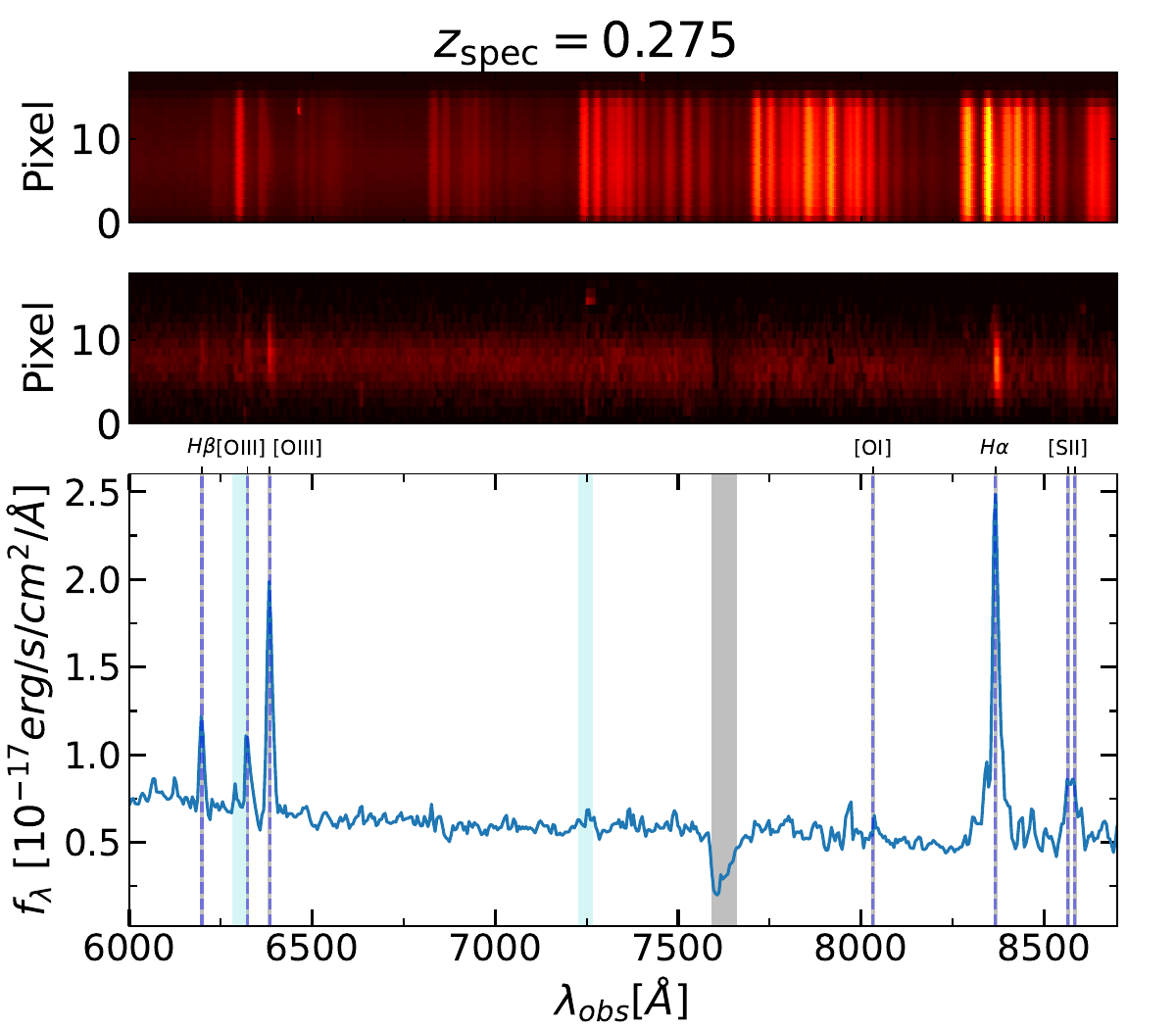}
    \end{center}
    \caption{Example of the sky emission correction process for an object of the faint subset. (Top) Slice of the observed 2D raw spectrum with a 3-arcsec slit using OSIRIS in MOS mode. The intensity of the emission from the sky makes it impossible to appreciate the contribution of the observed object. (Mid) Result obtained by applying the sky subtraction described in the text. This allowed us to recover the emission of the observed object. (Bottom) Final 1D spectrum obtained for this object. The redshift obtained is $z_\mathrm{spec}=0.275$ using the observed emission lines.}
    \label{fig:sky_corr}
\end{figure}

This method of sky emission subtraction gives even better results than the sky subtraction obtained in the case of 10-arcsec slits, where the sky can be selected in the 2D spectra. With the \textit{ON-OFF-ON} strategy, the 2D spectra to be corrected for sky contribution and the 2D sky spectra used have the same shape because they come from the same slit with precisely the same characteristics (i.e.\ slit irregularities, CCD curvature, and differential refraction of the light). In Fig. \ref{fig:sky_corr}, we can see an example of the sky subtraction for a 3-arcsec-long slit. The top panel is a slice of the 2D raw spectrum, where the sky emission completely hides the emission from the observed object. The central panel shows the result after applying the previous sky emission subtraction. It can be seen how the emission of the object is now fully visible owing to the good correction applied. The lower panel shows the 1D spectrum for that object with really strong spectral lines. The spectroscopic redshifts of the objects are determined from the observed lines in the spectra. In this case, the redshift obtained is $z_\mathrm{spec}=0.275$.

Flux calibration is the last step when working with 2D spectra, just before obtaining the 1D spectra. For each OB, at least one standard star was observed to perform that task. Calibration was applied using the standard IRAF procedure.

\begin{figure*}
    \begin{center}
        \includegraphics[trim= 0.5cm 1.5cm 0.5cm 1cm,width=1\textwidth]{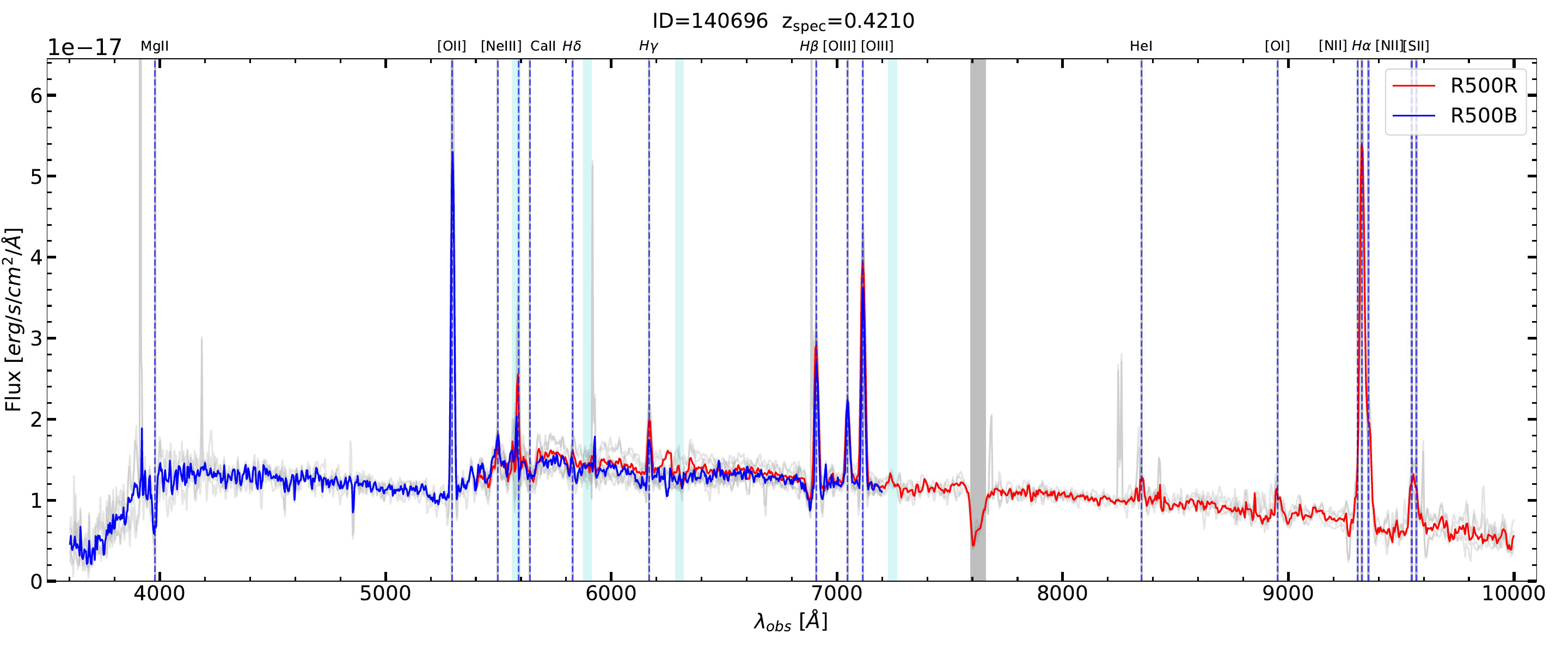}
    \end{center}
    \caption{Final 1D spectrum for an SFG in the faint subset at a spectroscopic redshift,  $z_\mathrm{spec}=0.421$. Blue and red lines are the final spectra for the R500B and R500R grisms, respectively. Individual observations with the grisms are plotted in grey, where some residuals of sky emission or cosmic ray detection are visible, but not in the final spectra owing to the application of the `average-sigma-clipping' algorithm. The vertical dashed blue lines mark the most prominent spectral features in this wavelength range at that redshift. The vertical cyan bands show the areas where sky emission is most important, and the wider grey band corresponds to the telluric absorption band.}
    \label{fig:1D_sample}
\end{figure*}

\subsubsection{1D spectra obtained}
Once the results of the previous data reduction steps were deemed satisfactory, we proceeded to obtain the final 1D spectra, which allowed us to know the nature of the observed objects and their main physical characteristics in the optical domain. 

As we have mentioned, each object was observed at least twice per OB. Furthermore, each mask was observed between three and four times with different grisms, as scheduled in Table \ref{tab:faint_camps}. Each object should finally have between three and eight observations per grism. The latter scenario is possible because an object could be observed more than once, just to fill in possible free spaces in the masks.

The desired outcome was a single 1D spectrum per grism. An `average-sigma-clipping' algorithm was applied to achieve this goal. The algorithm runs through each wavelength, discarding the flux points more than 2 $\sigma$ from the median and averaging the remaining points into a single point. The most common problems occurred when the object was observed in more than one mask and some of the observations were taken with a bright moon or a zero-order from a nearby slit with a bright target contaminating the spectrum. All these problems were managed using the average-sigma-clipping algorithm. In Fig. \ref{fig:1D_sample} we can see an example of the final 1D spectrum of an object at $z_\mathrm{spec}=0.421$, observed with the R500R and R500B grisms, represented by blue and red lines, respectively. The individual observations of this object with each grism, on which the average-sigma-clipping algorithm was applied, are plotted in grey. It can be seen how the algorithm has managed to correct for residuals from sky emission or cosmic ray detection.

\subsection{Bright subset reduction}
The reduction of the bright part was carried out using IRAF and the \texttt{hydra.dohydra}\footnote{\url{https://astro.uni-bonn.de/\~sysstw/lfa\_html/iraf/noao.imred.hydra.dohydra.htl\#h\_1}}package. This package was specifically developed for the reduction of data obtained with the HYDRA instrument. However, it allowed us to change its configuration and adapt it to the observations made with AF2. In this way, just by changing the parameters related to each instrument, this same tool was used for the different observing instruments of the bright part of our sample.

The \texttt{hydra.dohydra} task was used for scattered light subtraction, extraction, fibre throughput correction and wavelength calibration. It is a command language script that collects and combines the functions and parameters of many general-purpose tasks to provide a single complete data reduction path. The tool also allowed us to do a sky correction, but in this case, we only used the result of the combination of the fibres associated with the sky to obtain the average sky spectrum that we latter used as a final correction with the \texttt{Skycorr} \footnote{\url{https://www.eso.org/sci/software/pipelines/skytools/skycorr}} tool \citep{skycorr14}, which gave us better results in terms of the S$/$N quality of the final spectrum than those given by \texttt{dohydra}.

Before obtaining the final spectrum, the He and Ne lamp spectra were used for the wavelength calibration by employing an third-order Legendre polynomial for most of the objects in the range $0.03\leq$ RMS $\leq 0.07$. Flux calibration of fibre instruments is complicated to apply, as it depends directly on the quantum efficiency of each fibre at the time of measurement and its relationship with the others; that is, it is a function that depends directly on time and is internally variable fibre to fibre in the same way. Adding to this the fact that the observing routine of the bright subset was also very complicated, we decided that it was not necessary to apply this correction. Thus, these data are used to determine properties that do not require flux calibration such as spectroscopic redshifts, line widths, and flux ratios.

\begin{figure*}
    \begin{center}
        \includegraphics[trim= 0.5cm 1.5cm 0.5cm 1cm,width=1\textwidth]{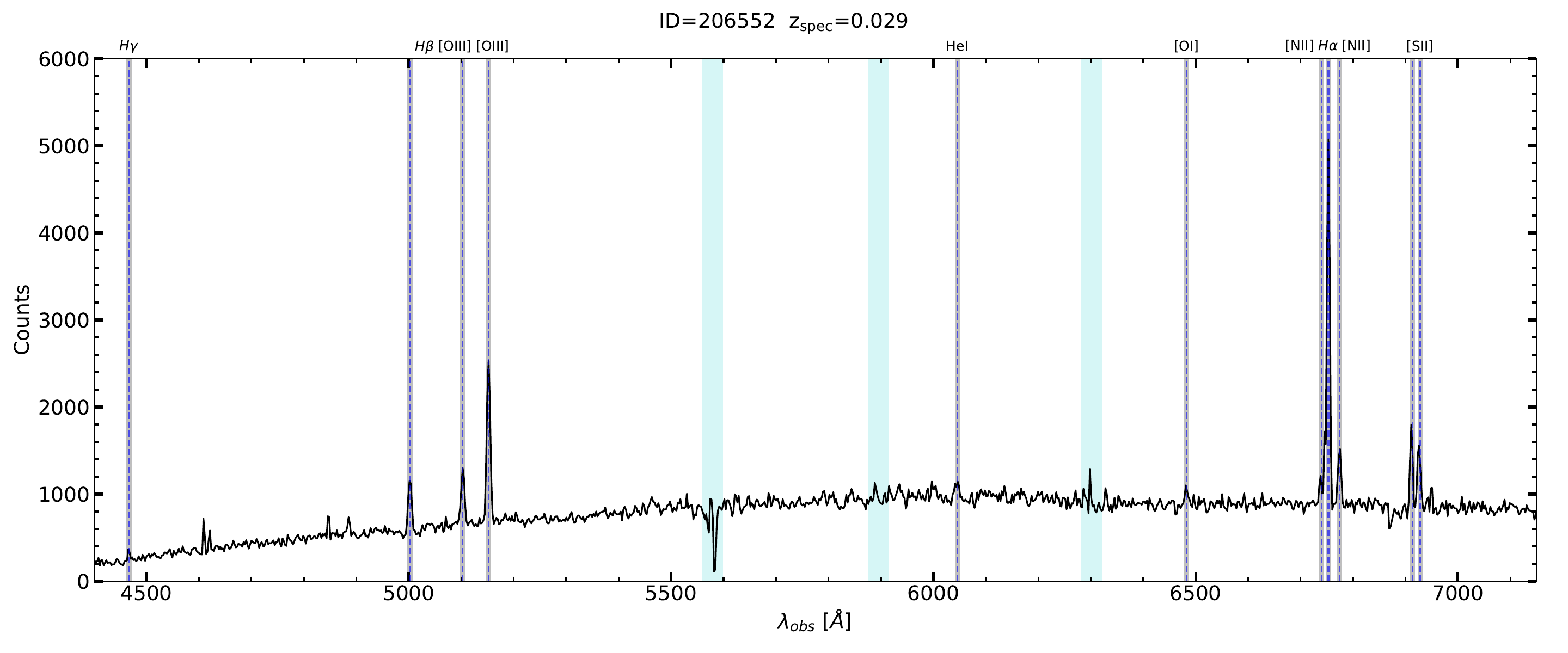}
    \end{center}
    \caption{Final 1D spectrum for a SFG in the bright subset at a spectroscopic redshift  $z_\mathrm{spec}=0.029$. This object was observed with WYIN/HYDRA using the R316R grism. The vertical cyan bands show the areas where sky emission predominates.}
    \label{fig:1D_sample_bri}
\end{figure*}

We likewise removed cosmic rays from individual images using the IRAF \texttt{lacos-spec} task \citep{lacospec01}. We obtained the average spectrum of each of the objects observed in the R600R and R300B networks for AF2-WYFFOS and the R316R network in HYDRA with the same reduction method. Figure \ref{fig:1D_sample_bri} shows an example of a 1D spectrum for a source in the bright subset observed with WYIN/HYDRA and the R316R grism.

%--------------------------------------------------------------------

\section{Lockman-SpReSO catalogue}\label{sec:6}
In this section, we present the first results obtained from the reduction of the Lockman-SpReSO data. Once the 1D spectra were acquired, the treatment was the same for objects in the bright part and those in the faint part, so the results shown in this section come from both subsets.

The whole process of obtaining a final spectrum per grism was executed automatically by the software infrastructure developed for this task described above. Although each component was tested and analysed, it was decided to conduct a visual inspection of all the results to look for possible errors in the process. Other important reasons for the visual inspection were to determine the spectroscopic redshifts of the objects by looking for the main spectral features, checking which spectra showed a stellar continuum and which did not, giving an initial quality flag of the spectral lines, and taking note of the objects with uncommon properties. In addition, further rounds of visual inspection were carried out after the fitting and measuring of the lines to make sure that everything worked correctly. The `Object cont.' column in Table \ref{tab:z_spec_category} shows the number of objects in which the stellar continuum was detected by the object category within Lockman-SpReSO.

The quality criterion imposed on the spectral lines in the visual inspection helped us to filter the objects according to that criterion. Possible values for the flag are: 1) no line appears in the spectrum; 2) the line has some error or is difficult to measure, or both (e.g.\ lines partially or totally under strong sky emission); 3) the line is weak but detectable; 4) the line is clearly visible with a moderate signal; and 5) an intense line with a high signal. This criterion defined the quality of the lines until we performed the line fits and obtained the equivalent widths (EWs) and S$/$N values.

\begin{figure*}
    \begin{center}
        \includegraphics[trim= 1cm 1cm 1cm 1cm,width=1\textwidth]{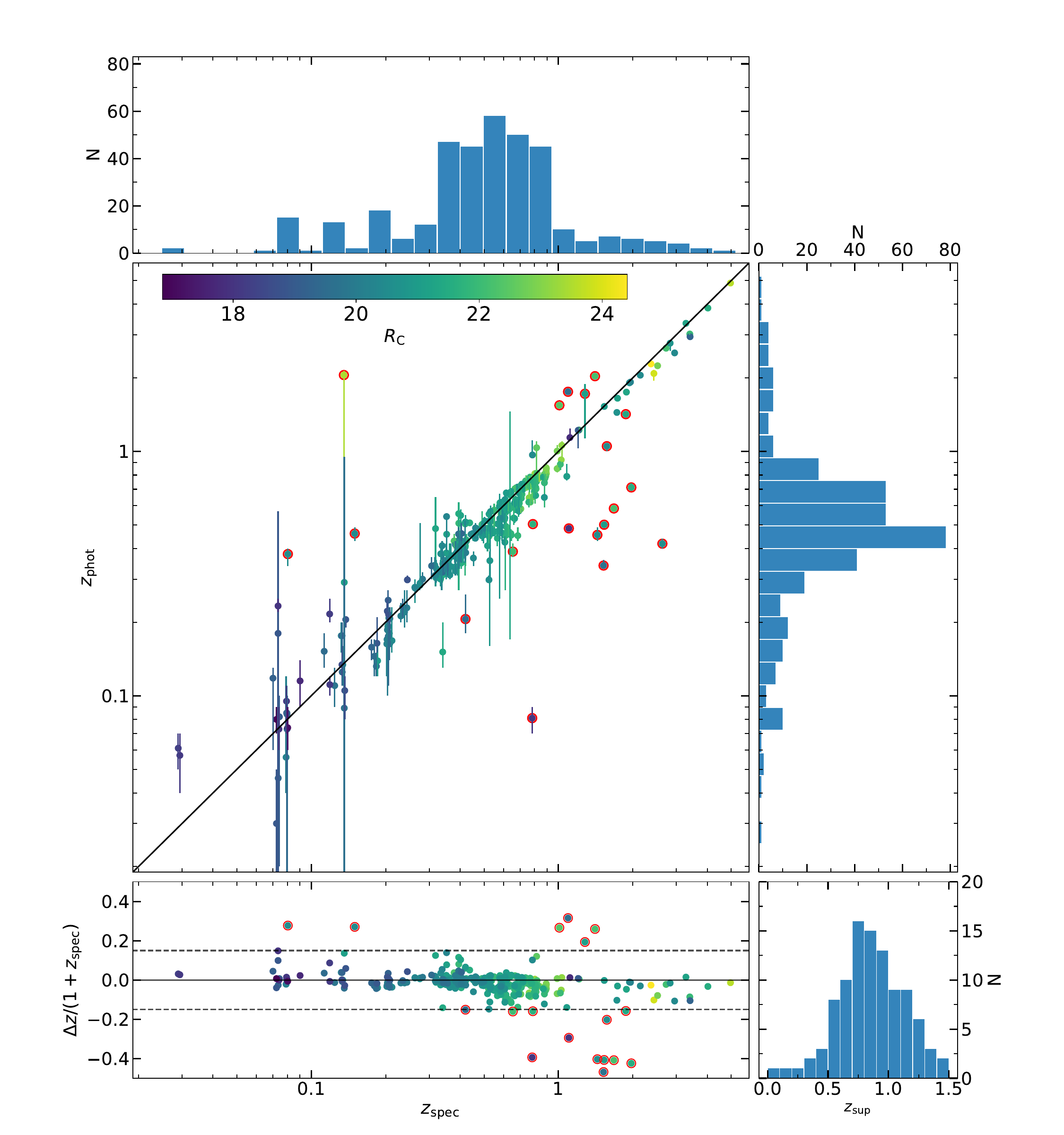}
    \end{center}
    \caption{Comparison between the photometric redshifts from \citetalias{Fotopoulou2012} and the spectroscopic redshifts measured in this study. The error bars in $z_\mathrm{phot}$ represent the range of 90$\%$ significance. The error bars in $z_\mathrm{spec}$ are not appreciable, and the median value is $3\times10^{-4}$. The colour code represents the $R_\mathrm{C}$ magnitude. In the bottom left panel, the difference between $z_\mathrm{phot}$ and $z_\mathrm{spec}$ divided by $z_\mathrm{spec}$ against $z_\mathrm{spec}$ is plotted. Dashed lines represent the limit in the outlier definition (see the Sect. \ref{sec:6.1}) and empty red circles represent the outliers. Because of the scale of the ordinate axis, there are points beyond the range of visualization. In the bottom right panel, the distribution of $z_\mathrm{sup}$ is shown.}
    \label{fig:zspec_vs_zphot}
\end{figure*}

\begin{table*}
    \centering
        \caption{Summary of the observed objects and the spectroscopic redshifts measured in this study, sorted by the preliminary categories described in Sect. \ref{sec:3.5}. The `Catalogue' column represents all objects in Lockman-SpReSO. The `Object cont.' column indicates how many objects per category have the stellar continuum detected in their spectra. The `$z_\mathrm{sup}$' column indicates the number of objects for which the redshift has been determined from a single spectral line}. It should be noted that an object could be observed in the faint and the bright subset so that the sum may be greater than the total. The same happens to the columns with the spectroscopic redshift.
        \label{tab:z_spec_category}
    \begin{tabular}{cc|r|rrr|r|rrr|r}
        \hline \hline
         CAT &  \begin{tabular}{@{}c@{}}Preliminary Object\\Type\end{tabular} &  Catalogue &   Observed &\begin{tabular}{@{}c@{}}Observed\\faint \end{tabular} &  \begin{tabular}{@{}c@{}}Observed\\bright\end{tabular} &  \begin{tabular}{@{}c@{}}Object\\cont.\end{tabular}  & $z_\mathrm{spec}$ & \begin{tabular}{@{}c@{}}$z_\mathrm{spec}$\\faint\end{tabular} &   \begin{tabular}{@{}c@{}}$z_\mathrm{spec}$\\bright\end{tabular} & $z_\mathrm{sup}$\\
        \hline
         1 & X-rayPoint + CatVarStars &         45 &      43 &           41 &              6 &               39 &      30 &     28 &     4 &  3\\
         2 &      High-Velocity Stars &         93 &      85 &           14 &             80 &               80 &       1 &      - &     1 &  -\\
         3 &           Radio Galaxies &         17 &      13 &           13 &              1 &                7 &       5 &      5 &     1 &  1\\
         4 &                      FIR &        902 &     838 &          772 &            106 &              503 &     305 &    258 &    76 & 90\\
         5 &                    1 + 4 &         12 &      11 &           11 &              1 &               11 &       7 &      7 &     1 &  3\\
         7 &                    3 + 4 &          4 &       4 &            3 &              1 &                4 &       3 &      2 &     1 &  1\\
        12 &                    1 + 2 &          1 &       1 &            1 &              1 &                1 &       1 &      1 &     - &  -\\
        20 &              RedQSOs(W4) &         23 &      21 &           21 &              - &                2 &       2 &      2 &     - &  -\\
        21 &           RedQSOs(FIRST) &          5 &       5 &            5 &              - &                1 &       - &      - &     - &  -\\
        23 &                   20 + 3 &          2 &       1 &            1 &              - &                1 &       - &      - &     - &  -\\
        24 &                   20 + 4 &         38 &      33 &           33 &              - &                5 &       1 &      1 &     - &  1\\
        25 &                   21 + 2 &          1 &       1 &            - &              1 &                1 &       1 &      - &     1 &  -\\
        26 &                   21 + 3 &          1 &       1 &            1 &              - &                1 &       1 &      1 &     - &  -\\
        \hline 
           &                    TOTAL &       1144 &    1057 &          916 &            197 &              656 &     357 &     305 &    85& 99\\
        \hline
    \end{tabular}
\end{table*}

\subsection{Redshift determinations}\label{sec:6.1}
To determine the spectroscopic redshifts, we imposed the condition that the spectrum should have at least two spectral features. The reason for doing so was to minimize, as much as possible, cases of false determinations due to ambiguous detection. It was also decided that, for objects with spectroscopic redshift in the literature, the presence of one line coinciding with the redshift would be the only condition. In total, 357 spectroscopic redshifts were obtained using both criteria and these are shown in Table \ref{tab:z_spec_category}, where they are split up by the object category and the subset to which they belong. 

We verified how many of these objects had a previous spectroscopic redshift determination. To do so, we used Table 5 of \citetalias{Fotopoulou2012}, which lists the works with spectroscopic studies that existed up to the date of publication for the LH field. In addition, we updated this information with those objects for which the SDSS survey provided redshifts, and finally we searched the NASA/IPAC Extragalactic Database\footnote{\url{https://ned.ipac.caltech.edu}} (NED) for the available information of the objects for which the redshift was determined in this work. A total of 89 objects already had their redshifts determined and coincided with those determined in this study. Hence, for 268 objects ($\sim$75\%), the spectroscopic redshift was determined for the first time.

Figure \ref{fig:zspec_vs_zphot} shows the comparison between the spectroscopic redshifts ($z_{\rm spec}$) measured in this study and the photometric redshifts ($z_{\rm phot}$) from \citetalias{Fotopoulou2012}. The error bars plotted for $z_{\rm phot}$ represent the range of 90 $\%$ significance reported by LePhare code (\citealt{Arnouts1999,Ilbert2006}). The errors in $z_{\rm spec}$ are also plotted, but are smaller than the data-point size, with a median value of $3\times10^{-4}$. The bottom panel shows the scatter of the difference between photometric and spectroscopic redshifts. The dashed grey lines represent the limit value defined by \cite{Hildebrandt2010} for flagging an outlier, defined by
\[
    \frac{|\Delta z|}{1+z_\mathrm{spec}} \geq 0.15,
\]
where \( |\Delta z| = |z_\mathrm{phot} - z_\mathrm{spec}|\). Applying this criterion, we found 4$\%$ of outliers in the spectroscopic redshift range $z_\mathrm{spec}<0.5$, 2$\%$ at $0.5 < z_\mathrm{spec} < 1.0$ and 34$\%$ for $z_\mathrm{spec}>1$. Outliers were marked with empty red circles in the Fig. \ref{fig:zspec_vs_zphot}. As expected, the determination of photometric redshifts is problematic for distant objects, even considering one out of three of the more distant objects as an outlier. Nevertheless, this result can be useful when using the photometric redshift to derive other quantities; for example, when the determination of the spectroscopic redshifts is not very clear because the spectral lines are not intense enough, or in cases where only one line is present in the spectrum, as we discuss below.

There are 105 objects in the LH-catalogue for which we detected only one emission-line feature in their spectra without any complementary information in the literature. An attempt was made to give a redshift value ($z_\mathrm{sup}$) based on the properties of the line found (intensity, observed wavelength, photometric redshift, object magnitudes, and credibility). This preliminary analysis made it possible to give a reliable redshift for 99 objects with only one spectral line. The bottom right panel of Fig. \ref{fig:zspec_vs_zphot} shows the distributions obtained for $z_{\rm sup}$, where $71\%$ of the objects have $z_\mathrm{sup}<1.0$ and the fraction of outliers in the photometric distribution are less than 3\%, so a redshift based on the line found in the spectra is very helpful. For all other objects, the photometric redshift was used with care, more weight being given to the other information available for that object. The last column in Table \ref{tab:z_spec_category} shows how the objects with $z_{\rm sup}$ are distributed among the different categories and their number. Figure \ref{fig:mag_Rc} shows the distribution of the $R_\mathrm{C}$ magnitude for the objects with $z_\mathrm{spec}$ (blue) and $z_\mathrm{sup}$ (red) values measured in this study.

In some cases, the determination of the spectroscopic redshift allowed us to clarify the nature of the objects under study. For example, for some of the candidate CV stars we, determined a range of values for $z_\mathrm{spec}$ (0.5263 $\leq$ $z_\mathrm{spec}$ $\leq$ 1.9387), which made it clear that they cannot be stars, but distant compact sources.

Although we already had $z_{\rm spec}$ measured, the final value for each object was calculated as a weighted mean of each of the redshifts obtained for each line after fitting. This is discussed further in the next section..

\begin{figure*}
    \begin{center}
        \includegraphics[width=1\textwidth]{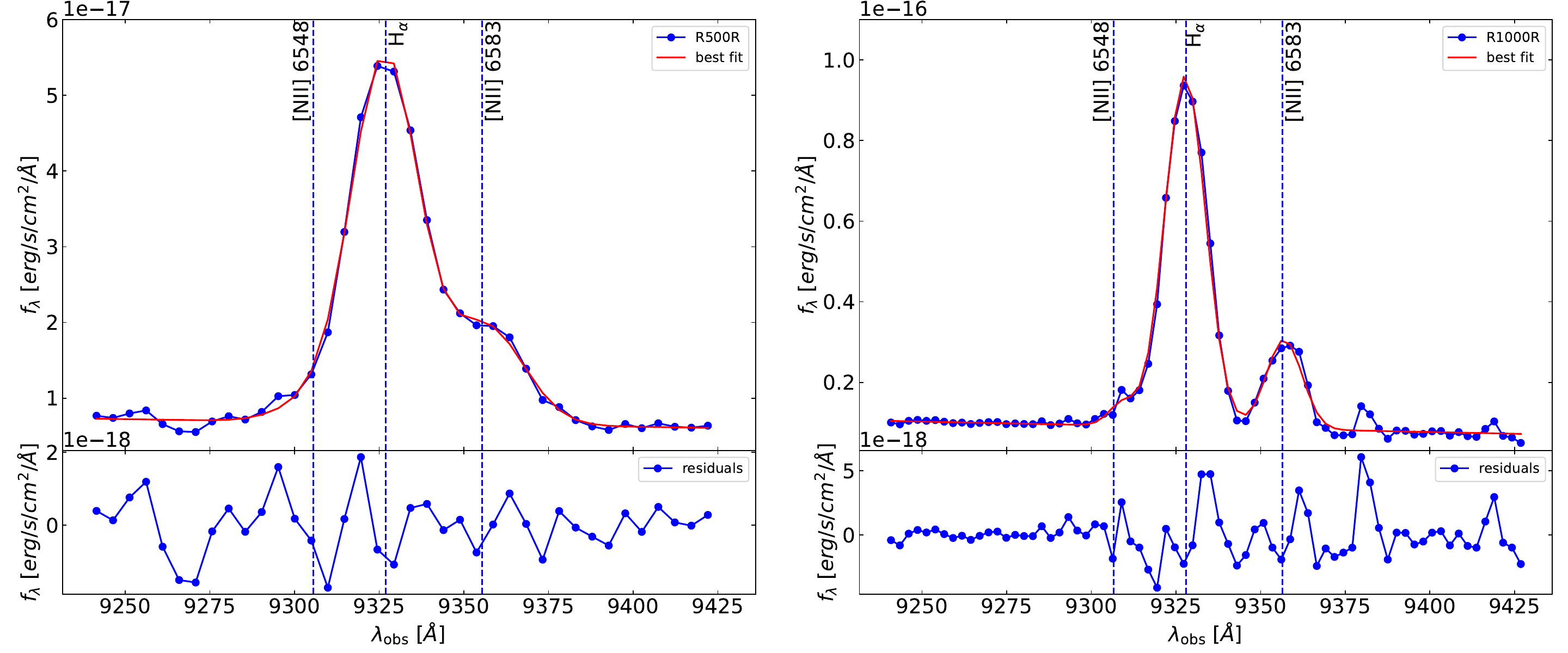}
    \end{center}
    \caption{Observed flux fitted to the same object of the faint subset with different grisms in the region of H$\alpha$ + [\ion{N}{II}] $\lambda$6548,6583. In the upper panels, blue dots and the blue line represent the observed flux, and the red line represents the best fit. The residuals of the fits are plotted in the bottom panels.}
    \label{fig:line_fit}
\end{figure*}

\begin{figure}
    \begin{center}
        \includegraphics[width=0.48\textwidth,trim= 2.5cm 2cm 1cm 1cm]{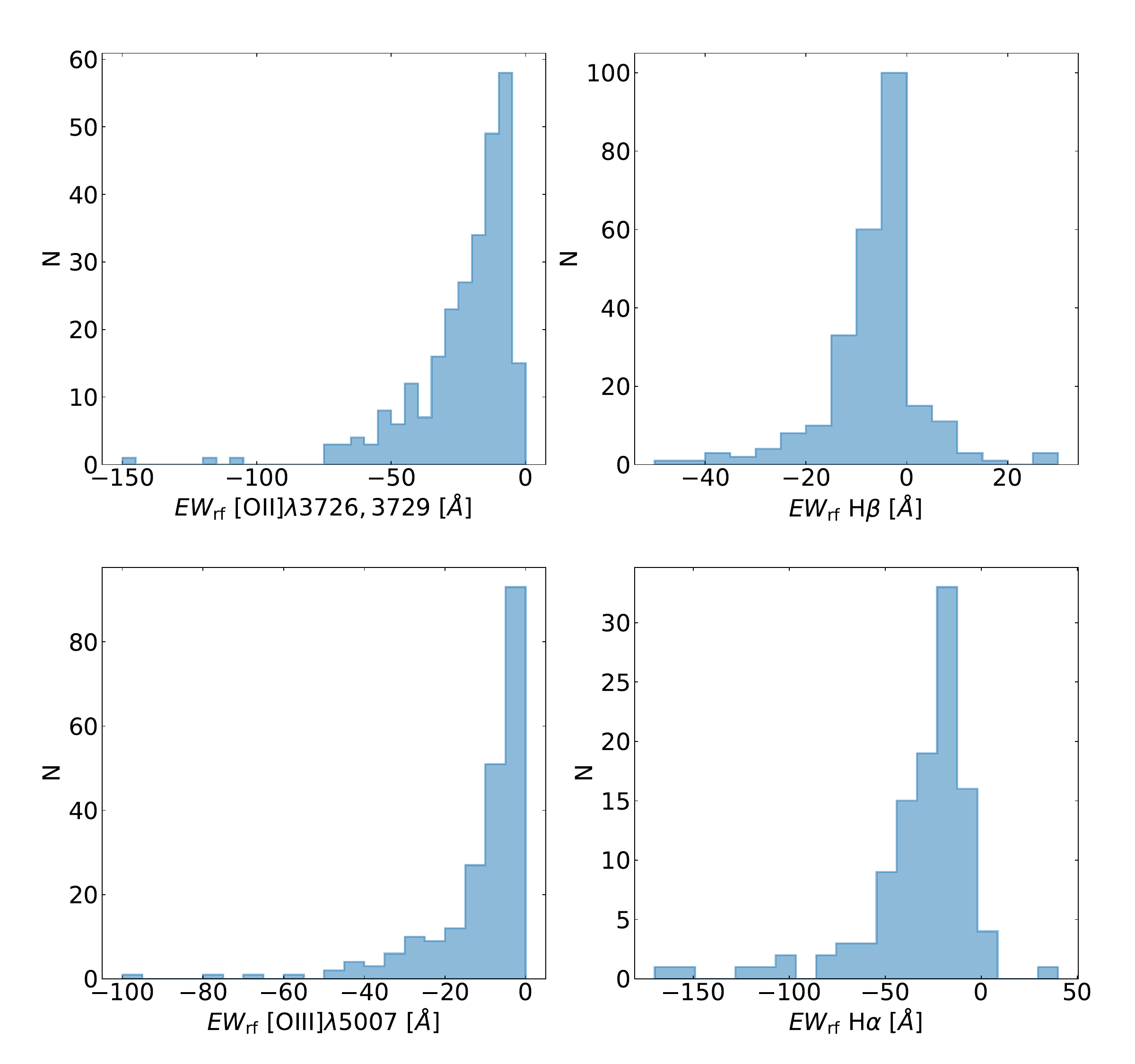}
    \end{center}
    \caption{Distribution of the rest-frame EWs for some of the most intense lines in the optical range for all the objects in the LH-catalogue with a measured spectroscopic redshift.}
    \label{fig:ew_observed}
\end{figure}

\subsection{Line measurement}\label{secc:6.2}
To measure the lines, we fitted them with a non-linear least-squares minimization routine implemented in Python  (\textit{LMFIT\footnote{\url{https://lmfit.github.io/lmfit-py/index.html}}}, \citealt{LMFIT2014}). Each line was fitted with a Gaussian profile plus a linear model to take into account possible continuum variations, all of which resulted in a total of five parameters to be determined: the position of the centre, the sigma, the amplitude for the Gaussian component, and the slope and intercept for the linear model. 

The parameters were set free, but initial values were needed. The initial values were adapted and certain constraints were imposed for minimizing the computing time for each fitted spectral line. For example, as we knew $z_{\rm spec}$, the centre of the line could be obtained and used as the initial value for the centre of the Gaussian component in the fitting process. One of the restrictions applied concerned the value for the amplitude of H$\beta$  when the H$\alpha$  line was available in the object's spectrum.  The lines were usually fitted from the most intense to the least intense so that the value of the H$\beta$ amplitude could not be greater than the calculated H$\alpha$ amplitude, thus restricting the space of values that the fitting programme had to explore. The same was applied to the other lines of the Balmer series and even for forbidden lines pairs such as [\ion{O}{III}] $\lambda$4959 \AA\ and [\ion{O}{III}] $\lambda$5007 \AA\ and [\ion{N}{II}] $\lambda$6548 \AA\ and [\ion{N}{II}] $\lambda$6583 \AA\, among others. Another constraint applied was related to the $\sigma$ of the Gaussians for the spectral lines that came from the same regions of the galaxy,such as the abovementioned forbidden line pairs. The initial $\sigma$ value of the Gaussian with which we fitted their line profiles should be almost identical because their nature is the same. Another important parameter that remained fixed during the fitting and was external to it, was the wavelength window used for each line fit. The size of this region depended on the line type, the redshift, and the object type. 

With the spectral resolution used in our observations, there were spectral lines very close in wavelength that could not be resolved because they were blended into a single feature. In these cases, we had to fit more than one Gaussian model at same time, one for each line plus a linear model. Even in cases where the resolution was sufficient to separate the components, they were so close that a separate fit was difficult to perform. One example is shown in Fig. \ref{fig:line_fit}, where the emission feature corresponds to a group of three spectral lines: H$\alpha$ + [\ion{N}{II}] $\lambda$6548,6583 \AA. In the left panel, the grism used to observe the objects is the R500R, which has approximately half the resolution of the grism used on the right (the R1000R grism). In both cases, the blue dots are the observed flux, the red line is the best fit, and the bottom panels represent the residuals of both fits. It is important to note that, when observed with the lowest resolution, the components were blended into the same feature but that, when performing a fit, the three components were well recovered. In the right panel, the improvement of resolution allowed us to separate the [\ion{N}{II}] $\lambda$6583 \AA\ line from the H$\alpha$ line, but not enough to perform independent fits. Another case in which multiple Gaussians were fitted simultaneously concerned objects with broad lines. As usual in these situations, we fitted one Gaussian for the narrow component and another for the broad component.

Figure \ref{fig:ew_observed} shows the rest-frame EWs obtained for some of the strongest lines in the optical range. By definition, emission lines have a negative EW, and absorption lines have a positive EW. For this reason, only the H$\alpha$ and H$\beta$ lines have positive values in Fig. \ref{fig:ew_observed} because the forbidden lines [\ion{O}{II}] $\lambda$3726,3729 \AA\ and [\ion{O}{III}] $\lambda$5007 \AA\ cannot be in absorption. Finally, in Table \ref{tab:sample_line_catalogue} a sample of the information available in the database is presented. Principal information for both the FT and PEP-catalogues is included.

\begin{sidewaystable*}
\caption{Sample of the information in Lockman-SpReSO}
\label{tab:sample_line_catalogue}
\begin{tabular}{ccccccccccccccccc}
\hline\hline
    ID &        RA &      DEC &  CAT &    mag & mag err&       FIR & FIR err &  zphot &     z     & z err     & z type & line     &z line   &  z line err   &   \\
 units &     (deg) &    (deg) &      &        &        &     (mJy) &  (mJy)  &        &           &           &        &          &         &               &   \\
 (1)   &     (2)   &    (3)   &  (4) &  (5)   &  (6)   &     (7)   &  (8)    &  (9)   &    (10)   & (11)      & (12)   & (13)     & (14)    & (15)          &   \\
\hline
 84957 & 163.07966 & 57.36440 &    4 & 19.085 &  0.002 & 260        &  30    &  0.082 & 0.07464   & 0.00006   & spec & [\ion{O}{III}] $\lambda$4959  & 0.0746   &       0.0002  &   \\
       &           &          &      &        &        &            &        &        &           &           &      & [\ion{O}{III}] $\lambda$5007  & 0.0747   &       0.0001  &   \\
       &           &          &      &        &        &            &        &        &           &           &      & [\ion{S}{II}] $\lambda$6716   & 0.0746   &       0.0002  &   \\
       &           &          &      &        &        &            &        &        &           &           &      & [\ion{S}{II}] $\lambda$6731   & 0.0746   &       0.0002  &   \\
128229 & 163.16754 & 57.61459 &    4 & 20.854 &  0.004 & 730        &  20    &  0.474 & 0.482938  &  0.000009 & spec & \ion{Ca}{II} $\lambda$3934    & 0.48505  &            -  &   \\
       &           &          &      &        &        &            &        &        &           &           &      & \ion{Ca}{II} $\lambda$3968    & 0.48368  &            -  &   \\
       &           &          &      &        &        &            &        &        &           &           &      & \ion{H}{$\beta$}              & 0.48305  &       0.00002 &   \\
       &           &          &      &        &        &            &        &        &           &           &      & \ion{H}{$\alpha$}             & 0.48291  &       0.00001 &   \\
\hline 
\end{tabular}

\bigskip \bigskip
\renewcommand\thetable{5} 
\caption{\textit{continued}}
\begin{tabular}{ccccccccccc}

\hline\hline
   ID  &   line obs. centre &    line obs. centre err &   line flux &  line flux err &  FWHM line  &  line FWHM err &  line EW &  line S$/$N &  line quality \\
 units &       (\AA)        &          (\AA) &   (erg s$^{-1}$ cm$^2$ $\times$ 10$^{-17}$)  & (erg s$^{-1}$ cm$^2$ $\times$ 10$^{-17}$) & (\AA) & (\AA) &(\AA) &(\AA)  &            \\
 (1)   &   (16)             &    (17)        &   (18) &  (19) &  (20)  &  (21) &  (22) &  (23) &  (24) \\
\hline
 84957 &      5329    &               1 & 17           &     3  &   21     &       3 &    -2.54 &     12 &     5  & \\
       &      5381.5  &             0.5 & 29           &     2  &   19     &       1 &    -7.03 &     19 &     5  & \\
       &      7217    &               1 & 15           &     2  &   17     &       2 &    -5.21 &     13 &     5  & \\
       &      7234    &               1 & 16           &     2  &   17     &       2 &    -5.75 &     14 &     5  & \\
128229 &      5842    &               - &  2           &    -   &    2     &      -  &     1.32 &      1 &     1  & \\
       &      5887    &               - &  2           &    -   &    3     &      -  &     1.85 &      1 &     1  & \\
       &      7209.5  &            0.1  & 38.2         &   0.9  &   10.0   &     0.2 &   -55.67 &     19 &     5  & \\
       &      9732.36 &           0.07  & 91           &     1  &   16.4   &     0.2 &  -241.34 &     35 &     5  & \\
\hline 

\end{tabular}
\tablefoot{Column (1) is the unique identification number for each object in the catalogue. Column (2) and (3) give the optical coordinates (J2000) of the object. Coordinates of counterparts in other bands, available in \citetalias{Fotopoulou2012}, and FIR coordinates are also included. Column (4) is the flag to indicate to which catalogue the object belongs: 1) X-ray point sources or CV star candidates;2) high-velocity halo stars; 3) poorly studied radio galaxies; 4) FIR-objects; 5) objects in both $CAT = 1$ and $CAT = 4$; 7) objects in both $CAT=3$ and $CAT=4$; 12) objects in both $CAT=1$ and $CAT=2$; 20) very red QSOs selected using the \cite{Glikman2013} method; 21) very red QSOs selected using the \cite{Ross2015} criteria; 23) objects in both $CAT=20$ and $CAT=3$; 24) objects in both $CAT=20$ and $CAT=4$; 25) objects in both $CAT=21$ and $CAT=2$; and 26) objects in both $CAT=21$ and $CAT=3$. Column (5) and (6) are the AB magnitudes and errors for al the bands available in \citetalias{Fotopoulou2012} from FUV (\textit{GALEX}) to 8 $\mu$m (\textit{Spitzer}; see Section 3 in \citetalias{Fotopoulou2012} for more details). Columns (7) and (8) give the FIR fluxes and errors at 24, 100, and 160 $\mu$m from the PEP work expressed in mJy \citep{Lutz2011}. Column (9) is the photometric redshift calculated in \citetalias{Fotopoulou2012}. Columns (10) and (11) are the object redshift and the error measured in this work. Column (12) is the flag to mark how the redshift has been obtained. Two values are possible:`spec' and `sup'. Columns (13)--(24) are the main parameters of the fitted line using the model explained in the text. Each parameter of the Gaussian and line model is included in the table, as well as the EW and S/N obtained in the fitting process, and the quality criterion assigned in the first visual inspection.}
\end{sidewaystable*}

\subsection{Stellar-mass and IR luminosity distributions}

In order to further characterize the scope of Lockman-SpReSO, basic parameters such as stellar mass and luminosity are essential. One of the most common ways to obtain them is by using SED fits to derive the physical properties of the objects from best-fit models. In the work of \cite{Shirley2019}, the authors performed a unification of the fields studied by \textit{Herschel} (1270 deg$^2$ and 170 million objects) and produced a general catalogue (HELP). In addition, they carried out SED-fitting studies on this catalogue \citep{Malek2018} to determine the properties of the objects using a previously determined photometric redshift \citep{Duncan2018}. 

Spectroscopic redshift gives us the advantage that, for a chosen cosmology, we know the distance of the object, which is very important for SED fitting and accurate translation of rest-frame models to the observed wavelength. Thus, a SED-fitting process was carried out to take advantage of the good photometric coverage collected in \citetalias{Fotopoulou2012}, plus the FIR information at 24, 100 and 160 $\rm \mu$m, together with the spectroscopic redshift determined in this study. 

As explained in Sect. \ref{sec:2}, the \citet[hereafter K21]{Kondapally2021} multi-wavelength catalogue includes recent observations of the LH field in the optical range. The SpARCS and RCSLenS surveys observed the LH field using both the CFHT/MegaCam instrument and the broad-band filters $u$, $g$, $r$, $i$, and $z$, also included in the \citetalias{Fotopoulou2012} catalogue from observations made with SDSS. The SpARCS and RCSLenS bands were therefore added to the Lockman-SpReSO catalogue by cross-matching both catalogues. The process was restricted to a maximum distance of 1.5 arcsec, the same as that used when we merged the PEP-catalogue with the FT- and OSIRIS-catalogues. It was found that 97\% of the objects had a counterpart in \citetalias{Kondapally2021} at an angular distance of less than 1.5 arcsec. Another reason for the fusion of catalogues was the completeness of the \citetalias{Kondapally2021} sample with respect to the Lockman-SpReSO catalogue; in other words, for the  $u$, $g$, $r$, $i$, and $z$ bands, we have information for 30\%, 38\%,40\%, 40\%, and 39\% of the objects, respectively, within SDSS observations from \citetalias{Fotopoulou2012}. In SpARCS and RCSLenS, we have information for 97\%, 97\%, 97\%, 86\%, and 94\% of the objects. This effect is also present within the \textit{GALEX} data in the FUV and NUV bands, with information for 13\% and 27\% of the objects, respectively, in the \citetalias{Fotopoulou2012} catalogue, and in \citetalias{Kondapally2021} we have information for 74\% and 74\% of the objects, respectively.

\begin{figure}
    \begin{center}
        \includegraphics[width=0.48\textwidth]{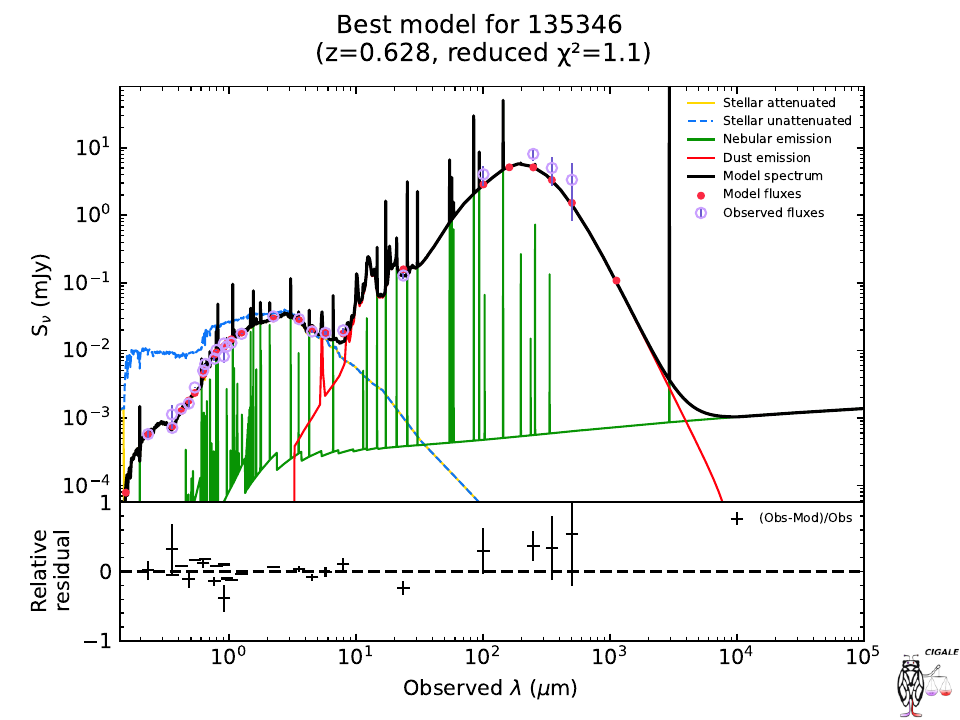}
    \end{center}
    \caption{Example of the best SED fit from CIGALE for a galaxy with z$_\mathrm{spec}=0.628$ measured in this work. The best model is plotted as a solid black line, the photometric information of the object is plotted as empty violet circles, and the red filled circles are the fluxes obtained by the best model. The individual contributions of the models used are also plotted where the yellow line represents the attenuated stellar component, the blue dashed line is the unattenuated stellar component, the red line is the dust emission, and the green line illustrates the nebular emission. The relative residuals of the flux for the best model are plotted at the bottom.}
    \label{fig:SED_fit}
\end{figure}

\begin{figure*}
    \begin{center}
        \includegraphics[width=1\textwidth, trim= 0cm 1cm 0cm 0cm]{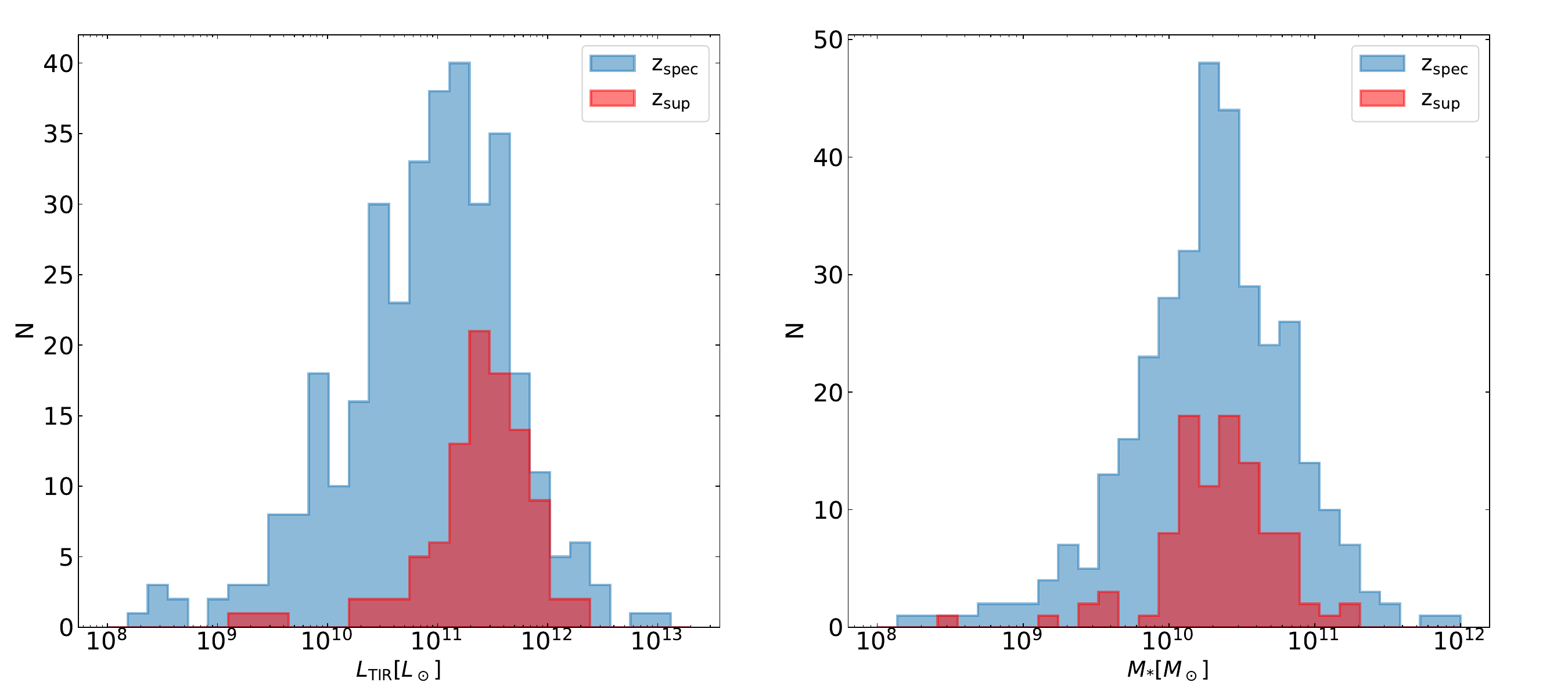}
    \end{center}
    \caption{Distribution of the total IR luminosity of the objects in the LH-catalogue derived from the SED fit using CIGALE (\textit{left}). Objects with $z_\mathrm{spec}$ are in blue, and the sample with $z_\mathrm{sup}$ is in red. The distribution of the stellar mass for the same objects was also derived from SED fitting (\textit{right}).}
    \label{fig:LTIR_Mass}
\end{figure*}

To try to cover the FIR range as well as possible, the Lockman-SpReSO catalogue was cross-matched with the HELP catalogue in order to get the flux in the 250, 350, and 500 $\rm \mu$m \textit{Herschel}/SPIRE bands. These photometric points helped to model the IR emission on the red side of the peak for the vast majority of objects in our spectroscopic sample. We found that 98.9\% of our objects have a HELP counterpart at an angular distance of less than 1.5 arcsec and 96.6\% have a HELP counterpart at less than 0.7 arcsec. The maximum separation was chosen to be 1.5 arcsec. Also, for the SMGs in the LH-catalogue, the JCMT/AzTEC 1.1 mm band value of \cite{Michalowski2012} was taken into account. Table \ref{tab:filters_c} compiles all the filters used in the SED fittings.

Therefore, for objects in the LH-catalogue with a calculated spectroscopic redshift, we fitted their SEDs using the CIGALE software (Code Investigating GALaxy Emission, \citealt{Bugarella2005}, \citealt{Boquien2019}). This code allowed us to perform SED fits from the UV to the radio (X-ray regimes are included in the latest update; we do not, however, use them in this study, \citealt{Yan2020}, \citeyear{Yan2022}). The code is based on the assumption of energy balance, that is to say that UV, optical and NIR light attenuated by dust is re-emitted in the redder ranges of the IR region. For each of the components involved in the SED fitting, CIGALE allowed us to choose between different models incorporated in the software. CIGALE performs a minimization of the $\chi^2$ statistic to select the best-fitting model. In addition, CIGALE also performs a Bayesian-based study to obtain a probability distribution of the physical parameters derived from the SED fit using all the models and their errors. Thus, using the available photometric and the spectroscopic redshift, we were able to obtain essential information about the objects in the LH-catalogue. 

For the stellar component, we used the \texttt{sfhdelayed} model, a star formation history (SFH) model with a nearly linear growth until a certain time ($\tau$), after which it drops smoothly, plus an exponential starburst. This model allowed us to fit both late-type galaxies (small $\tau$) and early-type galaxies (large $\tau$; \citealt{Boquien2019}). As in our study we are working with mostly infrared-emitting galaxies, we adopted a recent starburst to model the young population of stars ( i.e.\ SFGs). The intrinsic component of the stars was computed using the library of \cite[][model \texttt{bc03}]{Bruzual2003} with infra- and supra-solar metallicities and the IMF of \cite{Chabrier2003}. Nebular emission was also taken into account by adding the model to the CIGALE calculation. Extinction was incorporated using the \texttt{dustatt\_modified\_starburst} model, which is based on the extinction law of \cite{Calzetti2000}, to which the curve of \cite{Leitherer2002} between the Lyman break and 150 nm was added; in addition, both the slope and the UV bump could be modified. For the dust contribution, we used the \cite{Dale2014} templates (\texttt{dale2014} model) based on nearby SFGs that also added an AGN component. More sophisticated AGN models were not used because the purpose of these SED fits was the characterization of the Lockman-SpReSO sample as a whole, while assuming the AGN fraction-free parameter provided by the models of \citep{Dale2014} were adequate. Future studies of the Lockman-SpReSO project will carry out detailed investigations on object classification, and more individualized SED fits will study  each type of object more precisely. A summary of the selected models and the set of values used for the parameters of the models are given in Table \ref{tab:CIGALE}. An example of the SED-fitting process is shown in Fig. \ref{fig:SED_fit}, where the best fit obtained for a source in the Lockman-SpReSO catalogue is plotted together with the contribution of each of the models used. The lower part of Fig. \ref{fig:SED_fit} shows the relative residuals of the flux obtained in the fit.

Most of our objects were selected for their emission in the \textit{Herschel} bands (i.e.\ they are FIR emitters). Therefore, one of the first parameters to determine and describe the sample is the total IR luminosity ($L_\mathrm{TIR}$). There are different methods for obtaining the IR luminosity. Some of them derive $L_\mathrm{TIR}$ by using a monochromatic proxy \citep{Chary2001}, and others, as in the work developed by \cite[see also the references therein]{Galametz2013}, the IR luminosity is obtained from an analytic expression based on \textit{Spitzer} and \textit{Herschel} data. These studies are usually designed to use the rest-frame bands, which means that when we study distant galaxies, the bands are redshifted, and a correction is needed.

To surmount this difficulty, as we had the SED fits available, we could integrate the luminosity of the best fit in the IR range (usually between 8 and 1\,000 $\rm{\mu}$m) and directly obtain the $L_\mathrm{TIR}$. CIGALE already returns this luminosity and its error both from the best-fitting model and from the Bayesian approach. The left panel of Fig. \ref{fig:LTIR_Mass} shows the IR luminosity distribution obtained for the $z_\mathrm{spec}$ sample in blue and for the $z_\mathrm{sup}$ sample in red. We find that most of our objects (55$\%$) are in the LIRG regime (\(L_\mathrm{TIR} > 10^{11}L_\odot\)), 6$\%$ are ULIRGs  (\(L_\mathrm{TIR} > 10^{12}L_\odot\)), and less than 1$\%$ are hyper-luminous infrared galaxies (HLIRGs,\(L_\mathrm{TIR} > 10^{13}L_\odot\)).

Another parameter of importance for the general description of the sample is the stellar mass ($M_*$) of the objects. As for the IR luminosity, we used the value derived from the SED fit with CIGALE. Figure \ref{fig:LTIR_Mass} (right panel) shows the mass distribution obtained for the $z_\mathrm{spec}$ objects  in blue and for the $z_\mathrm{sup}$sample in red. Considering both distributions, 75$\%$ of the objects have stellar masses greater than $\log(M_*/M_\odot)=9.93$, 25$\%$ of them have a stellar mass greater $\log(M_*/M_\odot)=10.54$, and the median value is $\log(M_*/M_\odot)=10.28$.

\begin{figure}
    \begin{center}
        \includegraphics[width=0.48\textwidth, trim= 0.5cm 2cm 0.5cm 0cm]{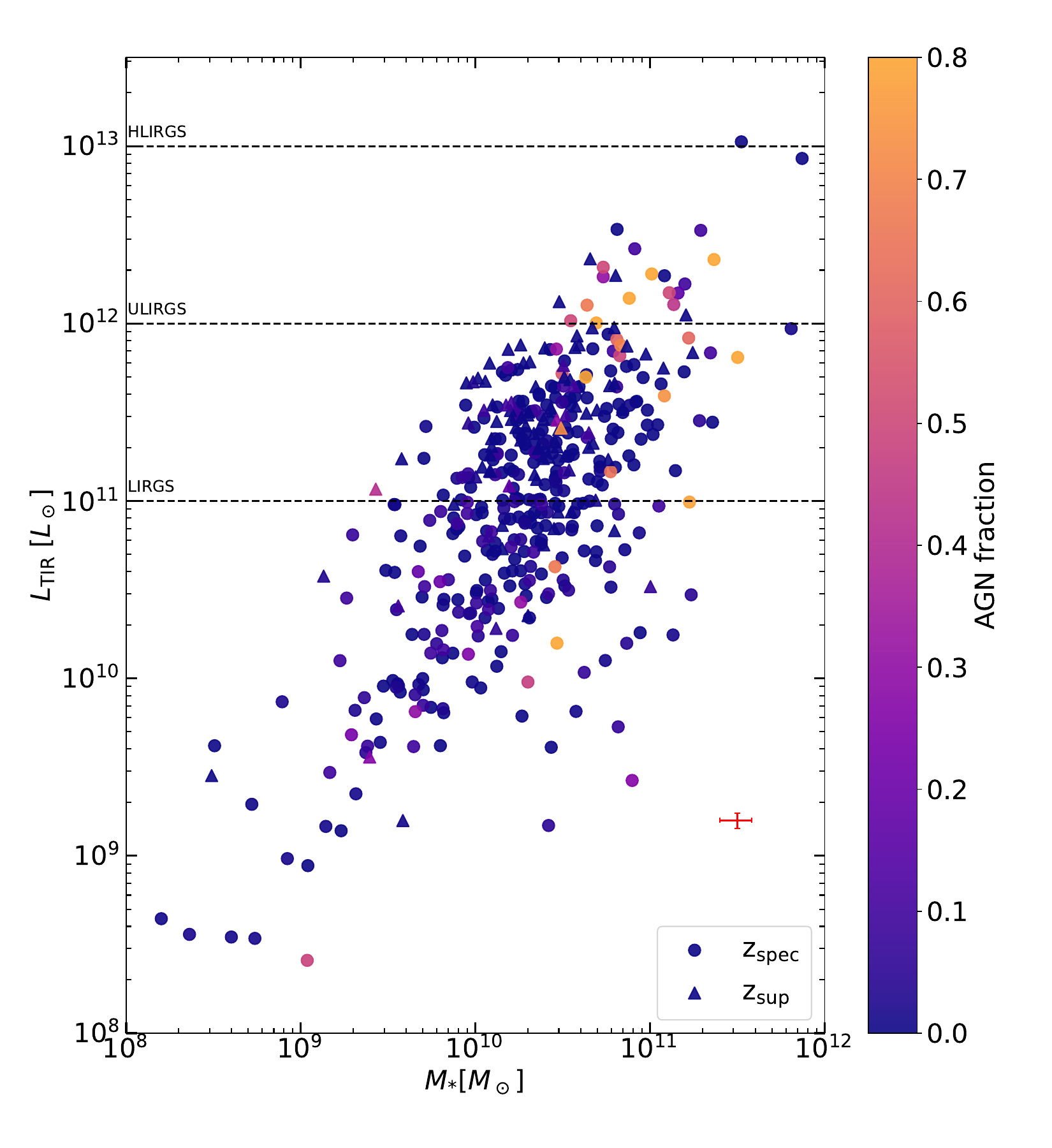}
    \end{center}
    \caption{Representation of the stellar mass and IR luminosity, both obtained from the CIGALE fits. The circles indicate the objects with $z_\mathrm{spec}$ calculated in this work, and the triangles represent the $z_\mathrm{sup}$ objects.  The median error on both axes is shown in red. The colour code indicates the fraction of AGN we obtained from the SED fits using the IR templates of \cite{Dale2014}. The boundary lines on which the LIRG, ULIRG, and HLIRG sources are defined are highlighted.}
    \label{fig:LTIR_Mass_fracAGN}
\end{figure}

Figure \ref{fig:LTIR_Mass_fracAGN} plots $M_*$ versus $L_\mathrm{TIR}$ obtained from the SED fittings by colour coding the AGN fraction derived from the IR templates used \citep{Dale2014}. It can be seen that the AGN fraction increases for the most luminous galaxies, which in turn are the most massive. This would be in agreement with the findings of \cite{Veilleux1995}, who showed that the probability of the source of ionization being due to nuclear activity increases with IR luminosity.

The information studied in this section will be used in forthcoming papers of the Lockman-SpReSO series. Stellar mass, IR luminosity, line measurements, and SED fits will play a key role in studying the physical properties of both FIR sources and secondary catalogue objects.

%--------------------------------------------------------------------

\section{Summary and timeline}
This paper presents the Lockman-SpReSO project and focuses on its scientific motivation, target selection, observational design and the first results from the reduction of the spectra. Lockman-SpReSO was created to try to fill the notable lack of spectroscopic information in one of the best large-scale fields, the Lockman Hole field, for high depth studies owing to its low hydrogen column density. In this way, spectroscopic redshifts and the main physical parameters such as gas metallicity, extinctions, SFR, and even SED fits can be studied for a specific selection of targets.

The Lockman-SpReSO catalogue contains 1144 sources of various kinds. All Lockman-SpReSO observations were carried out from  2014 to 2018, and all data have been satisfactorily reduced. The spectroscopic data obtained have been analysed and have produced spectroscopic redshifts for a total of 357 objects, from $z_\mathrm{spec} = 0.0290$ to $z_\mathrm{spec} = 4.9671$. For 99 objects with only one characteristic feature in their spectra, an attempt was  made to determine the redshift and resulted in a redshift range from $z_\mathrm{sup}=0.0973$ to $z_\mathrm{sup}=1.4470$. Furthermore, for those objects whose redshift was determined, only $\sim 25\%$ have a spectroscopic redshift available in the literature. Finally, the spectral lines of the objects were measured in order to establish the initial database of Lockman-SpReSO.

We performed an SED-fitting process using the spectroscopic redshift and CIGALE software, taking advantage of the wide photometric spectral coverage from the FUV to the FIR. The fit results allowed us to derive the IR luminosities and the stellar masses of the sources. Based on the $L_\mathrm{TIR}$ values derived from the SED fitting, about 55\% of the objects are  LIRGs, 6\% are ULIRGs and less than 1\% are HLIRGs. The stellar mass distribution has a minimum and a maximum value of $\log(M_*/M_\odot)=7.65$ and $\log(M_*/M_\odot)=12.07$, respectively, with a median value of $\log(M_*/M_\odot)=10.28$.

A data release is expected for late 2022 or early 2023, where the results of the type classification, extinction, gas metallicity and SFR determinations for the objects on which we have sufficient information will be presented. Studies will also be carried out in order to analyse the different classes of sources in the secondary catalogues. We shall try to perform observations in the NIR using other facilities for the objects with known redshifts that have emission lines in that wavelength range.

\begin{acknowledgements}
We thank the anonymous referee for their useful report. This work was supported by the Evolution of Galaxies project, of references AYA2017-88007-C3-1-P, AYA2017-88007-C3-2-P, AYA2018-RTI-096188-BI00, PID2019-107408GB-C41, PID2019-106027GB-C41, PID2021-122544NB-C41, and MDM-2017-0737 (Unidad de Excelencia María de Maeztu, CAB), within the \textit{Programa estatal de fomento de la investigación científica y técnica de excelencia del Plan Estatal de Investigación Científica y Técnica y de Innovación (2013-2016)} of the Spanish Ministry of Science and Innovation/State Agency of Research MCIN/AEI/ 10.13039/501100011033 and by `ERDF A way of making Europe'. This article is based on observations made with the Gran Telescopio Canarias (GTC) at Roque de los Muchachos Observatory on the island of La Palma, with the Willian Herschel Telescope (WHT) at Roque de los Muchachos Observatory on the island of La Palma and on observations at Kitt Peak National Observatory, NSF's National Optical-Infrared Astronomy Research Laboratory (NOIRLab Prop. ID: 2018A-0056; PI: Gonz\'alez-Serrano, J.I.), which is operated by the Association of Universities for Research in Astronomy (AURA) under a cooperative agreement with the National Science Foundation. This research has made use of the NASA/IPAC Extragalactic Database (NED), which is funded by the National Aeronautics and Space Administration and operated by the California Institute of Technology. J.N. acknowledge the support of the National Science Centre, Poland through the SONATA BIS grant 2018/30/E/ST9/00208. EB and ICG acknowledge support from DGAPA-UNAM grant IN113320. MP acknowledges the support from the Space Science and Geospatial Institute under the Ethiopian Ministry of Innovation and Technology (MInT). EA and MP acknowledge the support from the State Agency for Research of the Spanish MCIU through the Center of Excellence Severo Ochoa award to the Instituto de Astrof\'isica de Andaluc\'ia (SEV-2017-0709). JAD acknowledges the support of the Universidad de La Laguna through the Proyecto de Internacionalización y Excelencia, Programa Tomás de Iriarte 2022. The authors thank Terry Mahoney (at the IAC's Scientific Editorial Service) for his substantial improvements of the manuscript.
\end{acknowledgements}

%--------------------------------------------------------------------
\bibliographystyle{aa}
\bibliography{biblio} 

\onecolumn
\begin{appendix} %First appendix

\section{Photometric information}
\begin{table*}[h!]
\centering
\caption{Photometric filters used in the SED-fitting process and the number of objects with information in each of them.}
\label{tab:filters_c}
\begin{tabular}{cccccccc}
\hline
Telescope                          & Instrument                      & \multicolumn{6}{c}{Filter}                                                     \\ \hline
\multirow{2}{*}{\textit{GALEX}}    & \multirow{2}{*}{\textit{GALEX}} & \multicolumn{3}{c}{FUV}                 & \multicolumn{3}{c}{NUV}              \\
                                   &                                 & \multicolumn{3}{c}{842}                 & \multicolumn{3}{c}{842}              \\ \hline
\multirow{4}{*}{LBT}               & \multirow{2}{*}{LBCB}           & \multicolumn{3}{c}{U}                   & \multicolumn{3}{c}{B}                \\
                                   &                                 & \multicolumn{3}{c}{1072}                & \multicolumn{3}{c}{1040}             \\ \cline{2-8} 
                                   & \multirow{2}{*}{LBCR}           & \multicolumn{2}{c}{V}    & \multicolumn{2}{c}{Y}    & \multicolumn{2}{c}{z}    \\
                                   &                                 & \multicolumn{2}{c}{1011} & \multicolumn{2}{c}{1011} & \multicolumn{2}{c}{1028} \\ \hline
\multirow{2}{*}{Subaru}            & \multirow{2}{*}{Suprime}        & \multicolumn{2}{c}{Rc}   & \multicolumn{2}{c}{Ic}   & \multicolumn{2}{c}{z}    \\
                                   &                                 & \multicolumn{2}{c}{1144} & \multicolumn{2}{c}{1081} & \multicolumn{2}{c}{1134} \\ \hline
\multirow{2}{*}{SLOAN}             & \multirow{2}{*}{SLOAN}          & \multicolumn{2}{c}{u}    & g            & r         & i           & z          \\
                                   &                                 & \multicolumn{2}{c}{350}  & 442          & 470       & 470         & 454        \\ \hline
\multirow{2}{*}{CFHT}              & \multirow{2}{*}{MegaCAM}        & \multicolumn{2}{c}{u}    & g            & r         & i           & z          \\
                                   &                                 & \multicolumn{2}{c}{1102} & 1118         & 1105      & 980         & 1069       \\ \hline
\multirow{2}{*}{UKIRT}             & \multirow{2}{*}{WFCAM}          & \multicolumn{3}{c}{J}                   & \multicolumn{3}{c}{K}                \\
                                   &                                 & \multicolumn{3}{c}{1106}                & \multicolumn{3}{c}{1107}             \\ \hline
\multirow{4}{*}{\textit{Spitzer}}  & \multirow{2}{*}{IRAC}           & \multicolumn{2}{c}{3.6 $\mu$m}  & 4.5 $\mu$m          & 5.8 $\mu$m       & \multicolumn{2}{c}{8 $\mu$m}    \\
                                   &                                 & \multicolumn{2}{c}{972}  & 931          & 952       & \multicolumn{2}{c}{909}  \\ \cline{2-8} 
                                   & \multirow{2}{*}{MIPS}           & \multicolumn{6}{c}{24 $\mu$m}                                                         \\
                                   &                                 & \multicolumn{6}{c}{956}                                                        \\ \hline
\multirow{4}{*}{\textit{Herschel}} & \multirow{2}{*}{PACS}           & \multicolumn{3}{c}{100 $\mu$m}                 & \multicolumn{3}{c}{160 $\mu$m}              \\
                                   &                                 & \multicolumn{3}{c}{760}                 & \multicolumn{3}{c}{606}              \\ \cline{2-8} 
                                   & \multirow{2}{*}{SPIRE}          & \multicolumn{2}{c}{250 $\mu$m}  & \multicolumn{2}{c}{350 $\mu$m}  & \multicolumn{2}{c}{500 $\mu$m}  \\
                                   &                                 & \multicolumn{2}{c}{889}  & \multicolumn{2}{c}{889}  & \multicolumn{2}{c}{889}  \\ \hline
\multirow{2}{*}{JCMT}              & \multirow{2}{*}{AzTEC}          & \multicolumn{6}{c}{1.1 mm}                                                        \\
                                   &                                 & \multicolumn{6}{c}{18}                                                         \\ \hline
\end{tabular}
\end{table*}

\section{CIGALE input parameters}

\begin{table*}[h!]
    \centering
    \caption{Schedule of the parameters and models used in the CIGALE SED fitting.}
    \label{tab:CIGALE}
    \begin{tabular}{c c c}
    \hline \hline
    Model used & Parameter information & Values\\
    \hline
    SFH:                        & e-folding time main population  & 250, 500, 1000, 2000, 4000, 6000, 8000 [Myrs] \\
    delayed SFH with            & age of the main population      & 250, 500, 1500, 4000, 8000, 10000      [Myrs]\\
    optional exponential burst  & e-folding time of late burst    & 25, 50     [Myrs] \\
                                & age of the late burst           & 10, 20, 50 [Myrs] \\
                                & mass fraction late burst        & 0.0, 0.01, 0.05   \\
    \hline
    SSP:                       &    IMF          &   \cite{Chabrier2003}                                   \\
    \cite{Bruzual2003}  &    metallicity  &  0.0001, 0.0004, 0.004, 0.008, 0.02, 0.05    \\
    \hline
    Dust attenuation:    &  E(B-V) lines  & 0, 0.1, 0.2, 0.3, ..., 2.4, 2.5 \\
    power law modified   &  E(B-V) factor (line to continuum)  & 0.44 \\
     \cite{Calzetti2000} &  UV bump wavelength                  & 217.5 [nm] \\
                                 &  UV bump FWHM                       & 35 \\
                                 &  UV bump amplitude  & 0, 1.5, 3 \\
                                 &  Power law modification slope  & -0.2, 0 \\
                                 &  $R_{\rm V}$  & 3.1 \\
    \hline
    Dust emission &         AGN fraction & 0, 0.1, 0.3, 0.5, 0.8, 0.9\\
    \cite{Dale2014} &       Alpha slope & 0.125, 0.625, 1.0, 1.25, 1.5, 1.75, 2.0, 2.5, 3.0, 3.5, 4.0 \\
    \hline

    \end{tabular}
\end{table*}

\end{appendix}

\end{document}